 \documentclass[final,3p,times]{elsarticle}

\usepackage{lineno}
\usepackage[colorlinks]{hyperref}
\usepackage{amsmath}
\usepackage{amssymb}
\usepackage{lineno}
\usepackage{listings}
\usepackage{color}
\usepackage{textcomp}
\usepackage{graphicx}
\usepackage{subfigure}
\usepackage[ruled]{algorithm2e}

\biboptions{sort&compress}

\begin{document}

\begin{frontmatter}

  \title{A meshless method for compressible flows with the HLLC 
    Riemann solver}

  \author[MMU]{Z.~H.~Ma\corref{cor1}} \ead{z.ma@mmu.ac.uk}
  \author[JYU]{H.~Wang}
  \author[MMU]{L.~Qian}

  \cortext[cor1]{Corresponding author. Tel: +44 (0)161-247-1574}

  \address[MMU]{Centre for Mathematical Modelling and Flow Analysis, 
    School of Computing, Mathematics and Digital Technology,\\
    Manchester Metropolitan University, Manchester M1 5GD, United Kingdom}
  \address[JYU]{Department of Mathematical Information Technology, University of
   Jyv\"askyl\"a, Jyv\"askyl\"a, Finland}

\begin{abstract}
  The HLLC Riemann solver, which resolves both the shock waves and contact
  discontinuities, is popular to the computational fluid dynamics community
  studying compressible flow problems with mesh methods. Although it was
  reported to be used in meshless methods, the crucial information and procedure
  to realise this scheme within the framework of meshless methods were not
  clarified fully.  Moreover, the capability of the meshless HLLC solver to deal
  with compressible liquid flows is not completely clear yet as very few related
  studies have been reported. Therefore, a comprehensive investigation of a
  dimensional non-split HLLC Riemann solver for the least-square meshless method
  is carried out in this study.  The stiffened gas equation of state is adopted
  to capacitate the proposed method to deal with single-phase gases and/or
  liquids effectively, whilst direct applying the perfect gas equation of state
  for compressible liquid flows might encounter great difficulties in
  correlating the state variables. The spatial derivatives of the Euler
  equations are computed by a least-square approximation and the flux terms are
  calculated by the HLLC scheme in a dimensional non-split pattern. Simulations
  of gas and liquid shock tubes, moving shock passing a cylinder, internal
  supersonic flows in channels and external transonic flows over aerofoils are
  accomplished.  The current approach is verified by extensive comparisons of
  the produced numerical outcomes with various available data such as the exact
  solutions, finite volume mesh method results, experimental measurements or
  other reference results.
\end{abstract}


\end{frontmatter}


\section{Introduction}
\label{sec:introduction}
For compressible flow problems in computational fluid dynamics (CFD), the
physical features such as shock waves and/or contact discontinuities, across
which the fluid density, velocity and/or pressure vary abruptly, often appear in
the solution. They are usually referred as Riemann problems (named after
Bernhard Riemann) in mathematics \cite{Book-1997-Toro}. Based on the
characteristics analysis, the exact solutions can be obtained through iterations
and they perform well for simple problems \cite{Book-1973-Lax,
  Book-2002-LeVeque}.  However, exact Riemann solvers are computational
expensive for two- and three-dimensional problems, for which thousands or
millions of mesh elements are used to discretise the whole flow domain
\cite{1995-Anderson-p}. Therefore, great efforts have been made in the past
several decades to develop the efficient approximate Riemann solvers.

For this purpose, Harten, Lax and van Leer proposed the HLL Riemann solver with
a two-wave model resolving three constant states \cite{Paper-1983-Harten-35}. To
restore the intermediate wave missing from the HLL solver, Toro et
al. \cite{Report-1992-Toro,Paper-1994-Toro-25} proposed the HLLC Riemann solver composing a
three-wave model to separate four averaged states. Consequently the contact
discontinuity was identified by the HLLC solver \cite{Report-1992-Toro}. Batten
et al. further modified the wavespeeds estimate of the scheme, assuming the
intermediate wave speed equals to the average normal velocity between the
intermediate left and right acoustic waves \cite{Paper-1997-Batten-1553,
  Paper-1997-Batten-38}. The HLLC scheme is a versatile approximate Riemann
solver and it has attractive features including exact resolution of isolated
contact and shock waves, positivity preservation and enforcement of entropy
condition \cite{Paper-1997-Batten-1553,Paper-2004-Luo-304}.  It is now one of
the most popular approximate Riemann solvers used by the researchers
investigating compressible flow problems with mesh methods
\cite{Book-1997-Toro,Paper-1997-Batten-38,Paper-2009-Hu-6572,
  Paper-2006-Johnsen-715,Conference-2004-Lagumbay,
  Paper-2004-Luo-304,Paper-1997-Ivings-645}. 
   These facts trigger off our intention to
extend the HLLC scheme from mesh methods to meshless methods.

Meshless methods are relatively new and currently they are not as mature as mesh based
methods including finite difference, finite element and finite volume.  Since
the important work of Batina in meshless methods for computational gas dynamics
\cite{1993-Batina-p, 1992-Batina-p}, many studies have been accomplished to
explore the advantages of these methods for simulating external aerodynamics
problems using the JST (Jameson-Schmidt-Turkey) scheme or upwind schemes
\cite{2005-Chen-p439,2009-Katz-p5237,2004-Kirshman-p119,2005-Koh-p246,
  2008-Ma-p1926,2001-Morinishi-p551,2003-Sridar-p1,2007-Tota-p,2010-Wang-p98,
  2010-Hashemi-p2,2009-Ortega-p937,2002-Lohner-p1765,Paper-2011-Munikrishna-118}.
Some concerns were raised by researchers regarding the efficiency of meshless
methods, because these kinds of methods are usually not faster or even slower
than mesh methods on a per-point basis as indicated by
Batina\cite{1992-Batina-p}. While this topic is beyond the scope of the current
work, readers may refer to the following important works using implicit method
\cite{2001-Morinishi-p551, 2005-Chen-p439, 2010-Hashemi-p2}, adaptive method
\cite{2009-Ortega-p937, 2008-Ma-p1926}, hybrid method \cite{2004-Kirshman-p119,
  2006-Ma-p286} and multi-level cloud method \cite{2009-Katz-p5237}, which have
been accomplished to address this issue.  It was noticed by the current authors
that the HLLC scheme was claimed to be used in meshless methods for gas dynamics
problems \cite{2006-Luo-p618}. Unfortunately, it was not clearly stated whether
the flux terms were computed in a dimensional split or non-split manner, and the
wavespeeds estimate for the HLLC scheme was not clarified. After a relatively
thorough investigation of the above-mentioned works for meshless methods, it is
interesting to discover that these researcheres mainly focused on the
compressible gas flows and only the perfect gas equation of state (PG-EOS) was
adopted to correlate the state variables (density, pressure and
temperature/energy). Consequently, it is a little difficult to foresee whether
they will extend the meshless methods to the simulation of compressible liquids
for high pressure and/or high speed flow problems in hydrodynamics.

The objective of the present work is to conduct a comprehensive study of the
HLLC approximate Riemann solver within the framework of least-square meshless
method (LSMM) for compressible flows.  A dimensional non-split meshless HLLC
Riemann solver is presented with specific implementation details in this
paper. A corresponding step-by-step instructive computing algorithm, which is
easy to follow, is also provided in the paper.  The main work concentrates on
the compressible gas flows, which are very important for computational
aerodynamics. Referring the liquid flows in hydrodynamics, they are generally
considered to be incompressible due to their small density variations.  However,
for high pressure and/or high speed problems, simply assuming the liquids as
incompressible will encounter difficulties in numerical simulations
\cite{Paper-2009-Godderidge, Paper-2005-Murrone, Paper-2006-Johnsen,
  Paper-1998-Ivings-395, Paper-2009-Bredmose, Paper-2012-Plumerault,
  Paper-2013-Causon}. Therefore, a tentative study of the compressible liquid
flow problems with the meshless HLLC solver is also presented to gauge the
capability of the method. Considering the fact that the PG-EOS may be not very
suitable for liquids to correlate the state variables, the stiffened gas
equation of state (SG-EOS) is adopted in the present work. This enables the
presented method to effectively handle the gases and liquids with the
appropriate polytropic constants and pressure constants. To the best of our
knowledge, the implementation of the HLLC solver for the SG-EOS in meshless
methods has not been reported in the literature.

The rest of this paper is organised as follows. The governing equations
indicating the conservation of mass, momentum and energy for compressible flows
are presented in Section \ref{sec:governing-equations}. The basic theory of LSMM
is described in Section \ref{sec:basic-theory-lsmm-1}, the spatial
discretisation of the Euler equations is presented in Section
\ref{sec:meshless:discretization}. The HLLC approximate Riemann solver for LSMM
is illustrated in Section \ref{sec:numerical-method:flux} followed with the time
integration method described in Section \ref{sec:temp-discr}. Numerical examples
of one-dimensional gas and liquid shock tubes, two-dimensional moving shock
passing a cylinder, internal supersonic flows in channels and external transonic
flows over aerofoils are given in Section \ref{sec:test-cases} with intensive
comparisons to various available reference results. Conclusions are drawn in
Section \ref{sec:conclusions}.
\section{Governing equations}
\label{sec:governing-equations}

The present work focuses on the two-dimensional single-phase inviscid
compressible flows, of which the mathematical model is represented by the Euler
equations indicating the conservation of mass, momentum and energy. The
differential forms of these equations can be expressed as
\begin{equation} \label{eq:euler} \frac{\partial \mathbf{U}}{\partial t} +
  \frac{\partial \mathbf{E}}{\partial x} + \frac{\partial \mathbf{F}}{\partial
    y} = 0
\end{equation}
where $\mathbf{U}$ is a vector of conservative variables, $\mathbf{E}$ and
$\mathbf{F}$ are the flux terms, and are defined as
\begin{equation}
 \label{eq:wef}
 \mathbf{U}=\begin{bmatrix} \rho \\ \rho u \\ \rho v \\ \rho e_{\rm t} \end{bmatrix},\quad
 \mathbf{E}=\begin{bmatrix} \rho u \\ \rho u^2+p \\ \rho uv \\
   (\rho e_{\rm t}+p)u\end{bmatrix},\quad
 \mathbf{F}=\begin{bmatrix} \rho v \\ \rho uv \\ \rho v^2+p\\ (\rho e_{\rm t}+p)v \end{bmatrix}
\end{equation}
in which, $\rho$ is the density, $p$ is the pressure, $u$ and $v$ are the
components of velocity vector $\vec{V}$ along $x$ and $y$ axes respectively.
The total energy per volume $\rho e_{\rm t}$ is the sum of the internal energy
$\rho e_{\rm i}$ and
kinematic energy $\rho e_{\rm k}$, they are evaluated by the following formulae
\begin{subequations}
  \label{eq:total:energy}
  \begin{align}
    & \rho e_{\rm t}= \rho e_{\rm i} + \rho e_{\rm k} \\
    & \rho e_{\rm k}=\frac{1}{2}\rho(u^2+v^2) \\
    & \rho e_{\rm i}=e_{\rm i}(\rho,p)
  \end{align}
\end{subequations}
The stiffened gas equation of state (SG-EOS) is used in the present work, therefore the
internal energy is given by
\begin{equation}
  \label{eq:internal:energy}
  \rho e_{\rm i} = \frac{p+\gamma p_{\rm c}}{\gamma-1}
\end{equation}
where $p_{\rm c}$ is the pressure constant and $\gamma$ is the polytropic
constant.
For ideal gas, the pressure constant vanishes ($p_{\rm c}=0$) and the polytropic
constant equals to the ratio of specific heats ($\gamma=1.4$ for air).
The speed of sound can be calculated by the following formula
\begin{equation}
\label{eq:sound:speed}
c^{2}=\frac{\gamma}{\rho}(p+ p_{\rm c})
\end{equation}
\section{Numerical methods}
\label{sec:numerical-methods}

Meshless methods are relatively new for the compressible flow applications
compared to the traditional mesh-based methods including finite difference,
finite element and finite volume methods.  In this section, the basic theory of
LSMM is firstly described and the spatial discretisation procedure of the Euler
equations is explained. Then a dimensional non-split implementation of the HLLC
scheme in LSMM is presented. Finally the temporal discretisation of the
governing equations is given.

\subsection{Basic theory of LSMM}
\label{sec:basic-theory-lsmm-1}

\begin{figure}[h]
  \centering
  \includegraphics[height=5cm]{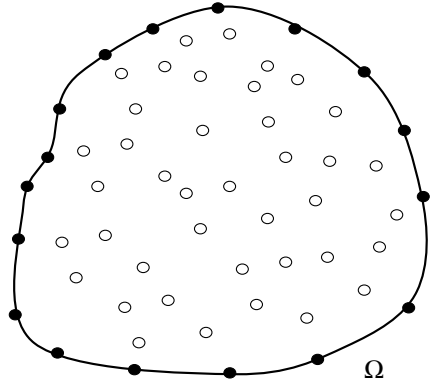}
  \caption{Meshless points distribution in the domain $\Omega$}
  \label{fig:domain:meshless}
\end{figure}

\begin{figure}[htp]
  \centering
  \includegraphics[height=4cm]{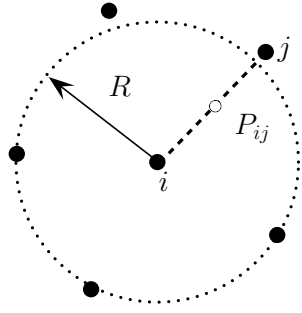}
  \caption{A typical meshless cloud of points}
  \label{fig:gridlesscloud}
\end{figure}

The essential idea of meshless methods is to introduce a number of scattered
points $P_i(i=1,2,3,\ldots,N)$ to a domain $\Omega$. The connectivity between
the points is not necessary to be considered. Fig. \ref{fig:domain:meshless}
gives an example of the domain discretisation using meshless points.  For each
point, several points around it are chosen to form a cloud of points
\cite{1993-Batina-p,1992-Batina-p,1996-Liu-p,1996-Onate-p3839}.
Fig. \ref{fig:gridlesscloud} shows a cloud of points $C_i$, in which the
point $i$ is named the centre and the other points are called
the satellites ($P_{ij}$ is the midpoint between $i$ and $j$).

The coordinate
difference between the satellite $j$ and the centre $i$ can be expressed as
\begin{equation}
  \label{eq:coor} h_j^i = x_{j} - x_{i},
  \qquad
  l_j^i = y_{j} - y_{i}
\end{equation}
For simplicity,  $h_j$ and $l_j$ are used to denote $h_j^i$ and $l_j^i$, respectively.
The vector ${\mathbf{r}}^i_j=(h_j, l_j)$ starts from $i$ to $j$,
its length is
\begin{equation}
  \label{eq:r}
  r^{i}_{j} = \sqrt{ h_j^2 + l_j^2  }
\end{equation}
and the reference radius of the cloud $C_i$ is defined as
\begin{equation}
  R_{i}=\max\left( r^{i}_{1},r^{i}_{2},\ldots,r^{i}_{j},\ldots,r^{i}_{M_i}\right)
\end{equation}
where $M_i$ is the total number of the satellites in the cloud.

Least square is adopted in the present work to approximate the spatial
derivatives of a function \cite{1993-Batina-p,1992-Batina-p,1996-Liu-p} and its
basic idea is described as follows.  Considering any 
 differentiable
function $f(x,y)$ in a given small domain $\Omega_i$, the Taylor series about a
point $P_i(x_i,y_i)$ can be expressed in the following form
\begin{equation}
  \label{eq:Taylor}
  f=f_i+a_1 h +a_2 l+ \frac{1}{2} a_3 h^2 + \frac{1}{2} a_4 l^2+
  a_5 hl +O(h^3,l^3)
\end{equation}
where $f=f(x,y)$ and $f_i=f(x_i,y_i)$ are the function values, $h=x-x_i$ and
$l=y-y_i$ are the coordinate differences between the points. The coefficients
$a_k (k=1,2,3,4,5)$ represent the partial derivatives of the function at
$P_i(x_i,y_i)$
\begin{equation}
  \label{eq:derivative}
  a_1=\left . \frac{\partial f}{\partial x} \right|_i,%
  ~ a_2=\left.\frac{\partial f}{\partial y}\right|_i, %
  ~ a_3=\left.\frac{\partial^2f}{\partial x^2}\right|_i, %
  ~ a_4=\left.\frac{\partial^2f}{\partial y^2}\right|_i, %
  ~ a_5=\left.\frac{\partial^2 f}{\partial x \partial y}\right|_i
\end{equation}
By keeping the terms until second order, the approximate function value at point
$P_j(x_j,y_j)$ is obtained
\begin{equation}
  \tilde{f}_j=f_i+a_1h_j+a_2l_j+a_3\frac{h^2_j}{2}
  +a_4\frac{l^2_j}{2}+a_5{h_jl_j}
\end{equation}
As the partial derivatives in the Euler equations are of first order, the terms
in the formula~\eqref{eq:Taylor} being kept to first order is usually
reasonable\cite{2009-Katz-p}. Consequently, the approximate value is
\begin{equation}
  \tilde{f}_j=f_i+a_1h_j+a_2l_j
\end{equation}
and the error between the exact and approximate values is
\begin{equation}
  e_j=f_j-\tilde{f}_j
  =f_j-\left(f_i+a_1h_j+a_2l_j \right)
\end{equation}
Then for cloud $C_i$, the total error can be estimated by the following norm
\begin{equation} \label{eq:error_norm} \Phi = \frac{1}{2}\sum^{M_i}_{j=1}e_j^2
\end{equation}

In order to minimise the error $\Phi$, its derivatives about $a_1$ and $a_2$
are set to zero
\begin{equation}
  \label{eq:least_square}
  \frac{\partial \Phi}{\partial a_1} = \frac{\partial
    \Phi}{\partial a_2} = 0
\end{equation}
therefore a set of linear equations is obtained 
\begin{equation}
  \label{eq:meshless:linear}
  \mathbf{Ax}=\mathbf{b}
\end{equation}
where
\begin{equation}
  \label{eq:meshless:linear:A}
  \mathbf{A}=
  \left[
    \begin{array}{ll}
      \sum h^2_j   &\sum h_j l_j  \cr 
      \sum h_j l_j &\sum l^2_j  \cr 
    \end{array}
  \right]
\end{equation}

\begin{equation}
  \label{eq:meshless:linear:xb}
  \mathbf{x}= \left[ \begin{array}{c}
      a_1 \\
      a_2 
    \end{array} \right] ~~~~~
  \mathbf{b}= \left[ \begin{array}{l}
      \sum h_j (f_j-f_i)  \\
      \sum l_j (f_j-f_i)  
    \end{array} \right]
\end{equation}
If the matrix $\mathbf{A}$ is not singular, this system of equations can be
solved by the following simple strategy
\begin{equation} \label{eq:inverse} \mathbf{x}=\mathbf{A}^{-1}\mathbf{b}
\end{equation}
The solutions can be written into a linear combinations of the function values at
different points
\begin{equation}
  a_1=\left . \frac{\partial f}{\partial x} \right |_i=\sum^{M_i}_{j=1} {\alpha}_j (f_j-f_i),~~ %
  a_2=\left . \frac{\partial f}{\partial y} \right |_i=\sum^{M_i}_{j=1} {\beta}_j (f_j-f_i)
\end{equation}
where ${\alpha}_j$ and ${\beta}_j$ are computed from Eq.
(\ref{eq:inverse}). They can also be estimated by the following formulae
\begin{equation}\label{eq:meshless:middle}
  a_1=\sum^{M_i}_{j=1} {\alpha}_{ij} (f_{ij}-f_i),~~
  a_2=\sum^{M_i}_{j=1} {\beta}_{ij} (f_{ij}-f_i)
\end{equation}
where the subscript $ij$ denotes the midpoint $P_{ij}$ between $i$ and $j$,
$f_{ij}$ is estimated at this midpoint
\cite{2002-Chen-p133,2005-Chen-p439,2001-Morinishi-p551}. The scalars at
$P_{ij}$ are two times of those at $P_{j}$
\begin{equation}
  \label{eq:meshless:scalar:midpoint}
  \alpha_{ij}=2\alpha_j, \qquad \beta_{ij}=2\beta_j
\end{equation}
For compressible flows, computing the spatial derivatives using
Eq.~\eqref{eq:meshless:middle} is suitable for the implementation of Riemann
solvers \cite{2008-Ma-p1926,2003-Sridar-p1}.  Some researchers introduce a
weight function to the error estimation formula~\eqref{eq:error_norm} and the
derived approach is named moving least square
\cite{2005-Chen-p439,2001-Morinishi-p551} or weighted least square
\cite{1996-Onate-p3839,2009-Ortega-p937,2006-Ma-p286}.

\subsection{Spatial discretisation }
\label{sec:meshless:discretization}

\begin{figure}[htp]
  \centering
  \includegraphics[width=0.4\textwidth]{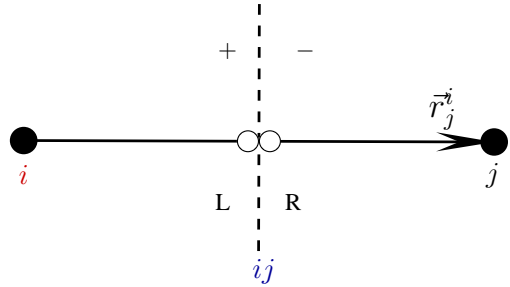}
  \caption{A satellite $j$, midpoint $ij$ and the centre $i$ in the cloud}
  \label{fig:midpoint}
\end{figure}

The formulae from \eqref{eq:meshless:linear} to \eqref{eq:meshless:middle}
provide a way to compute the spatial derivatives within a cloud of points. The Euler
equations are required to be satisfied for every cloud of points in the domain. 
For any cloud $C_i$, Eq. (\ref{eq:euler}) can be written as
\begin{equation}
\label{eq:meshless:euler:Pi}
  \left. \frac{\partial \mathbf{U}}{\partial t} \right\vert_{C_i} +
  \left(
    \frac{\partial \mathbf{E}}{\partial x} +
    \frac{\partial \mathbf{F}}{\partial y}
  \right)_{C_i} =0
\end{equation}
For simplicity,  the subscript $i$ is used to represent the cloud $C_i$ in the
following. Substitute Eq.~\eqref{eq:meshless:middle} into Eq. (\ref{eq:meshless:euler:Pi})
and it becomes
\begin{equation}
  \label{eq:euler:meshless}
  \frac{\partial \mathbf{U}_i}{\partial t} +
  \sum^{M_i}_{j=1}\left[ \left( \alpha_{ij} \mathbf{E}_{ij} + \beta_{ij}
      \mathbf{F}_{ij}\right) -\left( \alpha_{ij} \mathbf{E}_{i} + \beta_{ij}
      \mathbf{F}_{i} \right) \right]=0
\end{equation}
In order to calculate the flux terms in a non-split manner, the above equation
needs to be expressed in a compact form. For this purpose, a parameter $\lambda$
defined as
\begin{equation}
  \label{eq:lambda}
  \lambda=\sqrt{\alpha^2+\beta^2}
\end{equation}
and a 
vector $\vec{\eta}=(\eta_x,\eta_y)$ estimated by
\begin{equation}
  \label{eq:blending:vector}
  \eta_x=\frac{\alpha}{\lambda},\quad \eta_y=\frac{\beta}{\lambda}
\end{equation}
are introduced to Eq. \eqref{eq:euler:meshless}.  Now it can be written in the
following form
\begin{equation}
\label{eq:euler:centre:midpoint}
    \frac{\partial \mathbf{U}_i}{\partial t} +
  \sum^{M_i}_{j=1} ( \mathbf{G}_{ij} - \mathbf{G}_i ) \lambda_{ij}=0
\end{equation}
where
\begin{equation}
  \label{eq:G:function}
  \mathbf{{G}} = \eta_x \mathbf{E}  + \eta_y \mathbf{F}=
  \begin{bmatrix}
    \rho q \\
    \rho u q + p \eta_x \\
    \rho v q + p \eta_y \\
    (\rho e_{\rm t}+p) q
  \end{bmatrix}
\end{equation}
in which $q$ is the dot product of $\vec{V}$ and $\vec{\eta}$
\begin{equation}
  q= u \cdot \eta_x  +v \cdot \eta_y 
\end{equation}
Modern programming languages such as Fortran 90/95 provide advanced array
operations, which makes it easy for the computer codes be optimised on a machine
with vector processors. Based on these considerations, the flux term is advised
to be computed by the following vectorised formula
\begin{equation}
\label{eq:G:function:vector}
\mathbf{{G}} =q\mathbf{U}+p\mathbf{N}_q
\end{equation}
where $\mathbf{N}_q=[0,\eta_x,\eta_y,q]^{\rm T}$. %

As illustrated in Fig. \ref{fig:midpoint}, the flux function $\mathbf{G}$ at the
midpoint $P_{ij}$ is evaluated by
\begin{equation}
  \label{eq:G:midpoint}
  \mathbf{G}_{ij}=\mathbf{G}(\mathbf{U}_{ij}^{\rm L}, \mathbf{U}_{ij}^{\rm R})
\end{equation}
A simple method to calculate the conservative variables for the left side {\bf
  L} and the right side {\bf R} is
\begin{equation}
  \label{eq:wL:wR:1}
    \mathbf{U}_{ij}^{\rm L} = \mathbf{U}_i,\quad
    \mathbf{U}_{ij}^{\rm R} = \mathbf{U}_j
\end{equation}
This is the classical first-order Godunov scheme, which assumes a piecewise
constant distribution of the flow variables. To improve the accuracy, a
piecewise linear reconstruction of the data is adopted in this research. When
reconstructing the data, one option is to choose the conservative variables,
another is to reconstruct the characteristic variables and the other way is to
select the primitive variables. In the present work, the data is reconstructed
with the primitive variables $\mathbf{W}=(\rho,u,v,p)$ due to its simplicity
compared to the conservative and characteristic variables
\begin{subequations}
  \label{eq:data:reconstruction}
  \begin{align}
    & \mathbf{W}_{ij}^L = \mathbf{W}_i + \frac{1}{2} \nabla \mathbf{W}_i \cdot \mathbf{r}_j^i \\
    & \mathbf{W}_{ij}^R = \mathbf{W}_j - \frac{1}{2} \nabla \mathbf{W}_j \cdot \mathbf{r}_j^i
  \end{align}
\end{subequations}
As high order schemes tend to produce spurious oscillations in the vicinity of
large gradients \cite{Book-1997-Toro}, a slope or flux limiter needs to be used
to satisfy the TVD constrains \cite{Paper-1997-Harten-260}. In the present work,
the following slope limiter is employed 
\begin{subequations} \label{eq:limiter}
  \begin{align}
    \varphi^{\rm L}&=
    \frac{\nabla\mathbf{W}_i\cdot\vec{{r}}_j^i\Delta\mathbf{W}_j^i+
      \left| \nabla\mathbf{W}_i\cdot\vec{{r}}_j^i\Delta\mathbf{W}_j^i
      \right|+\epsilon}
    {(\nabla\mathbf{W}_i\cdot\vec{{r}}_j^i)^2+(\Delta\mathbf{W}_j^i)^2+\epsilon}\\
    \varphi^{\rm R}&=
    \frac{\nabla\mathbf{W}_k\cdot\vec{{r}}_j^i\Delta\mathbf{W}_j^i+
      \left| \nabla\mathbf{W}_k\cdot\vec{{r}}_j^i\Delta\mathbf{W}_j^i
      \right|+\epsilon}
    {(\nabla\mathbf{W}_k\cdot\vec{{r}}_j^i)^2+(\Delta\mathbf{W}_j^i)^2+\epsilon}
  \end{align}
\end{subequations}
where $\Delta \mathbf{W}_j^i = \mathbf{W}_j - \mathbf{W}_i$. The small value
$\epsilon$ is introduced to prevent null division in the smooth regions where
differences approach zero, it is set as $\epsilon=10^{-12}$ in this research. In
summary, the conservative variables are estimated by the reconstructed primitive
variables
\begin{subequations}
  \begin{align}
    & \mathbf{W}_{ij}^L = \mathbf{W}_i + \frac{1}{2}  \varphi^{\rm L} \nabla \mathbf{W}_i \cdot \mathbf{r}_j^i \\
    & \mathbf{W}_{ij}^R = \mathbf{W}_j - \frac{1}{2} \varphi^{\rm R} \nabla
    \mathbf{W}_j \cdot \mathbf{r}_j^i \\
    & \mathbf{U}_{ij}^{\rm L,R}=\mathbf{U}(\mathbf{W}_{ij}^{\rm L,R}) 
  \end{align}
\end{subequations}

\subsection{The HLLC approximate Riemann solver}
\label{sec:numerical-method:flux}

\begin{figure}[htp]
  \centering
  \includegraphics[width=0.5\textwidth]{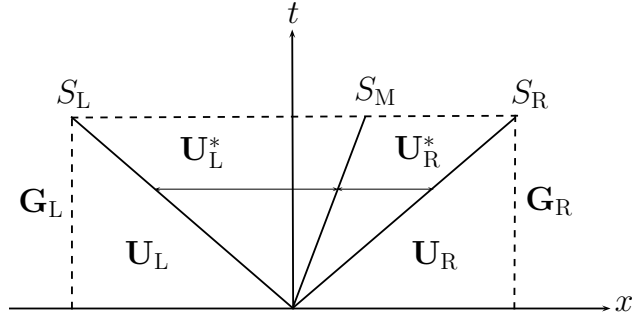}
  \caption{The HLLC Riemann solver}
  \label{fig:HLLC}
\end{figure}

As mentioned in Section \ref{sec:introduction}, the HLLC Riemann solver
introduces an intermediate contact wave to the original two-wave model of the
HLL Riemann solver. Consequently, there are two intermediate averaged states
($\mathbf{U}_{\rm L}^*$ and $\mathbf{U}_{\rm R}^*$) separated by the contact
wave $S_{\rm M}$ as shown in Figure \ref{fig:HLLC}. The approximate solution may
be expressed by four averaged states
\begin{equation}
  \mathbf{U}_{ij}=
  \left\{
    \begin{array}{ll}
      \mathbf{U}_{\rm L}    & 0 \le S_{\rm L}\\
      \mathbf{U}_{\rm L}^{*} & 0 \in (S_{\rm L}, S_{\rm M}]\\ 
      \mathbf{U}_{\rm R}^{*} & 0 \in (S_{\rm M}, S_{\rm R})\\
      \mathbf{U}_{\rm R}    & 0 \ge S_{\rm R}               
    \end{array}
  \right.
\end{equation}
The corresponding midpoint flux is given by
\begin{equation}
\label{eq:G:flux:0}
  \mathbf{G}_{ij}=
  \left\{
    \begin{array}{ll}
      \mathbf{G}_{\rm L}      & 0 \le S_{\rm L}\\              
      \mathbf{G}_{\rm L}^{*}   & 0 \in (S_{\rm L}, S_{\rm M}]\\
      \mathbf{G}_{\rm R}^{*}   & 0 \in (S_{\rm M}, S_{\rm R})\\
      \mathbf{G}_{\rm R}      & 0 \ge S_{\rm R}                
    \end{array}
  \right.
\end{equation}

Following the strategy of Toro \cite{Book-1997-Toro} by applying the
Rankine-Hugoniot conditions across the wave speeds $S_{\rm L}$ and $S_{\rm R}$,
the following important formulations are obtained
\begin{subequations}
  \label{eq:G:left:right:star:1}
  \begin{align}
  & \mathbf{G}_{\rm L}^* = \mathbf{G}_{\rm L} + S_{\rm L}^* (\mathbf{U}_{\rm L}^*
  - \mathbf{U}_{\rm L}) \label{eq:G:left:star:1}\\
  & \mathbf{G}_{\rm R}^* = \mathbf{G}_{\rm R} + S_{\rm R}^* (\mathbf{U}_{\rm R}^*
  - \mathbf{U}_{\rm R})\label{eq:G:right:star:1}
  \end{align}
\end{subequations}
In order to compute the intermediate state flux, the averaged state conservative
flow variables are necessary. To provide these information, 
the intermediate flux is assumed to be a function of the averaged state flow
variables as suggested by Batten et al. \cite{Paper-1997-Batten-1553,Paper-1997-Batten-38}
\begin{subequations}
  \label{eq:G:left:right:star:2}
  \begin{align}
  &\mathbf{G}_{\rm L}^*=q_{\rm L}^*  \mathbf{U}_{\rm L}^* + p_{\rm L}^* \mathbf{N}_{q_{\rm L}^*} \label{eq:G:left:star:2}\\
  &\mathbf{G}_{\rm R}^*=q_{\rm R}^*  \mathbf{U}_{\rm R}^* + p_{\rm R}^*
  \mathbf{N}_{q_{\rm R}^*} \label{eq:G:right:star:2}
  \end{align}
\end{subequations}
Substitute Eq.~(\ref{eq:G:left:right:star:1}) into the above equation, we obtain
\begin{subequations}
  \label{eq:G:left:right:star:3}
  \begin{align}
    &(q_{\rm L}^*-S_{\rm L}^*) \mathbf{U}_{\rm L}^* = \mathbf{G}_{\rm L} -
    S_{\rm L}^*\mathbf{U}_{\rm L}^* - p_{\rm L}^*\mathbf{N}_{q_{\rm L}^*} \\
    &(q_{\rm R}^*-S_{\rm R}^*) \mathbf{U}_{\rm R}^* = \mathbf{G}_{\rm R} -
    S_{\rm R}^*\mathbf{U}_{\rm R}^* - p_{\rm R}^*\mathbf{N}_{q_{\rm R}^*} 
  \end{align}
\end{subequations}
Notice that $\mathbf{G}_{\rm L}$ and $\mathbf{G}_{\rm R}$ can be estimated by Eq.
(\ref{eq:G:function:vector}) straightforwardly,
\begin{subequations}
  \label{eq:G:left:right:4}
  \begin{align}
  &\mathbf{G}_{\rm L} =q_{\rm L}   \mathbf{U}_{\rm L}  + p_{\rm L}
  \mathbf{N}_{q_{\rm L} } \label{eq:G:left:right:4:left} \\
  &\mathbf{G}_{\rm R} =q_{\rm R}   \mathbf{U}_{\rm R}  + p_{\rm R} 
  \mathbf{N}_{q_{\rm R} } \label{eq:G:left:right:4:right}
  \end{align}
\end{subequations}
Substitute Eq.~(\ref{eq:G:left:right:4}) to (\ref{eq:G:left:right:star:3}), we
obtain
\begin{equation}
  \label{eq:U:left:star}
  \mathbf{U}_{\rm L}^* = \frac{(q_{\rm L}-S_{\rm L})\mathbf{U}_{\rm L}+(p_{\rm
      L}\mathbf{N}_{q_{\rm L}}-p_{\rm L}^* \mathbf{N}_{q_{\rm L}^*})}{q_{\rm L}^* - S_{\rm L}}
\end{equation}
and 
\begin{equation}
  \label{eq:U:right:star}
  \mathbf{U}_{\rm R}^* = \frac{(q_{\rm R}-S_{\rm R})\mathbf{U}_{\rm R}+(p_{\rm
      R}\mathbf{N}_{q_{\rm R}}-p_{\rm R}^* \mathbf{N}_{q_{\rm R}^*})}{q_{\rm R}^* - S_{\rm R}}
\end{equation}
Once the intermediate velocity component $q_{\rm L,R}^*$ and pressure
$p_{\rm L,R}^*$ are known, the conservative variables can be easily obtained by
the above two equations. Based on the condition
\begin{equation}
  \label{eq:HLLC:pressure:condition}
  p_{\rm M}=p_{\rm L}^*=p_{\rm R}^*
\end{equation}
(there is no jump of pressure across the contact wave) and a convenient
assumption \cite{Paper-1997-Batten-1553} setting the contact wave speed to the
following
\begin{equation}
  \label{eq:HLLC:velocity:condtion}
  S_{\rm M}=q_{\rm L}^*=q_{\rm R}^*=q_{\rm M}
\end{equation}
the intermediate wave speed can be calculated by
\begin{equation}
  \label{eq:intermediate:wave:speed}
  S_{\rm M}=
  \frac{\rho_{\rm R}q_{\rm R}(S_{\rm R}-q_{\rm R})-\rho_{\rm L}q_{\rm L}(S_{\rm L}-q_{\rm L})+p_{\rm L}-p_{\rm R}}
  {\rho_{\rm r}(S_{\rm R}-q_{\rm R})-\rho_{\rm L}(S_{\rm L}-q_{\rm L})}
\end{equation}
and the intermediate pressure may be estimated as
\begin{equation}
  \label{eq:intermediate:pressure}
  p_{\rm M}=\rho_{\rm L}(q_{\rm L}-S_{\rm L})(q_{\rm L}-S_{\rm M})+p_{\rm L}
  =\rho_{\rm R}(q_{\rm R}-S_{\rm R})(q_{\rm R}-S_{\rm M})+p_{\rm R}
\end{equation}
The left and right states wave speeds are computed by
\begin{equation}
\label{eq:left:wave:speed}
  S_{\rm L}=\min(q_{\rm L}-c_{\rm L},\tilde{q}-\tilde{c})
\end{equation}
and 
\label{sec-1-1-6-2}%
\begin{equation}
  \label{eq:right:wave:speed}
  S_{\rm R}=\max(q_{\rm R}+c_{\rm R},\tilde{q}+\tilde{c})
\end{equation}
where $\tilde{q}$ is an averaged velocity component evaluated by
\begin{subequations}
  \label{eq:Roe:average:velocity}
  \begin{align}
    &\tilde{u}=\frac{\sqrt{\rho_{\rm L}} u_{\rm L}+\sqrt{\rho_{\rm R}} u_{\rm R}}{\sqrt{\rho_{\rm L}}+\sqrt{\rho_{\rm R}}}\\
    &\tilde{v}=\frac{\sqrt{\rho_{\rm L}} v_{\rm L}+\sqrt{\rho_{\rm R}} v_{\rm R}}{\sqrt{\rho_{\rm L}}+\sqrt{\rho_{\rm R}}}\\
    &\tilde{q}=\eta_x\tilde{u}+\eta_y\tilde{v}
  \end{align}
\end{subequations}

For ideal gas, the averaged speed of sound $\tilde{c}$ can be computed from the
averaged enthalpy $\tilde{h}$ as stated by Roe \cite{Paper-1981-Roe-357} and
Batten et al. \cite{Paper-1997-Batten-1553,Paper-1997-Batten-38}
\begin{subequations}
  \label{eq:sound:speed:average}
  \begin{align}
    &\tilde{h}=\frac{\sqrt{\rho_{\rm L}} h_{\rm L}+\sqrt{\rho_{\rm R}} h_{\rm R}}{\sqrt{\rho_{\rm L}}+\sqrt{\rho_{\rm R}}}\\
    &\tilde{c}=\sqrt{(\gamma-1)\left[\tilde{h}-\frac{1}{2}(\tilde{u}^2+\tilde{v}^2)\right]} 
  \end{align}
\end{subequations}
However, it is not clear whether this also stands for compressible liquids with
SG-EOS (\ref{eq:internal:energy}) as the corresponding formulations were not
provided in their papers. Therefore, effort is made here to remove the
ambiguity.  To clarify this point, the relation between the enthalpy and speed
of sound needs to be understood
\begin{equation}
  \label{eq:enthapy:1}
  \rho h=(\rho e_{\rm t}+p)
\end{equation}
Substitue Eq.~\eqref{eq:total:energy} and \eqref{eq:internal:energy} into
Eq. \eqref{eq:enthapy:1} then divided it by $\rho$, we obtain
\begin{subequations}
  \label{eq:enthapy:2}
  \begin{align}
  h&=(\rho e_{\rm t}+p)/\rho \\
   &=\frac{1}{\gamma-1}\frac{\gamma}{\rho}(p+p_{\rm c})+\frac{1}{2}(u^2+v^2) 
  \end{align}
\end{subequations}
Substitue Eq. \eqref{eq:sound:speed} into Eq. \eqref{eq:enthapy:2}, we get
\begin{equation}
  \label{eq:enthapy:3}
  h =\frac{1}{\gamma-1}c^2+\frac{1}{2}(u^2+v^2)     
\end{equation}
Therefore, the averaged speed of sound for compressible liquids can also be
appropriately recovered from the averaged enthalpy given by
Eq. \eqref{eq:sound:speed:average}.


Substitute Eq. (\ref{eq:HLLC:pressure:condition}),
(\ref{eq:HLLC:velocity:condtion}), (\ref{eq:intermediate:wave:speed}) and
(\ref{eq:intermediate:pressure}) to Eq. (\ref{eq:G:left:right:star:2}), the
intermediate left and right states flux can be expressed by the following forms
\begin{subequations}
  \begin{align}
   &   \mathbf{G}_{\rm L}^{*} = S_{\rm M} \mathbf{U}_{\rm L}^{*} +
   p_{\rm M}\mathbf{N}_{S_{\rm M}}\label{eq:G:left:star:3}\\
   &  \mathbf{G}_{\rm R}^{*} = S_{\rm M} \mathbf{U}_{\rm R}^{*} + p_{\rm
     M}\mathbf{N}_{S_{\rm M}} \label{eq:G:right:star:3}
  \end{align}
\end{subequations}
where $\mathbf{N}_{S_{\rm M}}= [ 0, \eta_x, \eta_y, S_{\rm M}]^{\rm T}$.  When
programming the practical code on a computer, we separate the process into two
steps. Firstly, we need to estimate the three wave speeds (the left wave, right
wave and intermediate contact wave). Then we compute the flux as Eq
(\ref{eq:G:flux:0}) according to the relations of these wave speeds.
The whole procedure to evaluate $\mathbf{G}_{ij}$ is summarised in Algorithm
\ref{sec:hllc-appr-riem}.

\begin{algorithm}[H]
\label{sec:hllc-appr-riem}
  \SetKwInOut{Input}{Input}
  \SetKwInOut{Output}{Output}
  \Input{$\mathbf{U}_{\rm L}, \mathbf{U}_{\rm R}$, $\vec{\eta}$}
  \Output{$\mathbf{G}$ }
  \Begin(Wave speeds estimate){
      $\mathbf{W}_{\rm L} \longleftarrow \mathbf{U}_{\rm L}$\;
      $\mathbf{W}_{\rm R} \longleftarrow \mathbf{U}_{\rm R}$\;
      $\tilde{q}, \tilde{c} \longleftarrow \mathbf{W}_{\rm L}, \mathbf{W}_{\rm
        R}$\;
      $S_{\rm L} \longleftarrow $ Eq.~(\ref{eq:left:wave:speed}) \;
      $S_{\rm R} \longleftarrow $ Eq.~(\ref{eq:right:wave:speed})\;
      $S_{\rm M} \longleftarrow $ Eq.~(\ref{eq:intermediate:wave:speed})\;
      $p_{\rm M} \longleftarrow $ Eq.~(\ref{eq:intermediate:pressure})\;    
    }
  \Begin(Flux determination){
      \uIf{$0<S_{\rm L}$}{$\mathbf{G}=\mathbf{G}_{\rm L}\longleftarrow$ Eq.~(\ref{eq:G:left:right:4:left})\;}
      \uElseIf{$S_{\rm L}\le 0 < S_{\rm M}$}{
        $\mathbf{U}_{\rm L}^* \longleftarrow$ Eq. (\ref{eq:U:left:star}) \;
        $\mathbf{G}=\mathbf{G}_{\rm L}^* \longleftarrow$ Eq.~(\ref{eq:G:left:star:3}) \;}
      \uElseIf{$S_{\rm M}\le 0 < S_{\rm R}$}{
        $\mathbf{U}_{\rm R}^* \longleftarrow$ Eq. (\ref{eq:U:right:star}) \;
        $\mathbf{G}=\mathbf{G}_{\rm R}^* \longleftarrow$ Eq.~(\ref{eq:G:right:star:3}) \;}
      \Else{$\mathbf{G}=\mathbf{G}_{\rm R}\longleftarrow$
        Eq.~(\ref{eq:G:left:right:4:right})\;}
    }
  \caption{Non-split meshless HLLC Riemann solver}
\end{algorithm}

\subsection{Temporal discretisation}
\label{sec:temp-discr}
 In the present work, the Euler equations are treated by the method-of-line,
 which separates the temporal and spatial spaces. The semi-discrete form of the
 governing equations is given by
 \begin{equation}
   \label{eq:euler:semi}
   \frac{{\rm d} \mathbf{U}}{{\rm d} t}=\mathbf{R}
 \end{equation}
 where $\mathbf{R}$ represents the residual. For cloud $i$, a forward difference
 discretisation of Eq.~(\ref{eq:euler:semi}) is
 \begin{equation}
   \frac{\mathbf{U}^{n+1}_i-\mathbf{U}^n_i}{\Delta t}=\mathbf{R}_i
 \end{equation}
An explicit four-stage Runge-Kutta scheme is applied to update the solution from time
level $n$ to $n+1$,
 \begin{subequations}
   \begin{align}
     &\mathbf{U}^{(0)}_{i} = \mathbf{U}^n_{i} \\
     &\mathbf{U}^{(1)}_i =\mathbf{U}^{(0)}_i+\alpha_1\Delta t_i\mathbf{R}^{(0)}_i \\
     &\mathbf{U}^{(2)}_i =\mathbf{U}^{(0)}_i+\alpha_2\Delta t_i\mathbf{R}^{(1)}_i \\
     &\mathbf{U}^{(3)}_i =\mathbf{U}^{(0)}_i+\alpha_3\Delta t_i\mathbf{R}^{(2)}_i \\
     &\mathbf{U}^{(4)}_i =\mathbf{U}^{(0)}_i+\alpha_4\Delta t_i\mathbf{R}^{(3)}_i \\
     &\mathbf{U}^{n+1}_{i} = \mathbf{U}^{(4)}_i
   \end{align}
 \end{subequations}
 where $\alpha_1=\tfrac{1}{4}$, $\alpha_2=\tfrac{1}{3}$,
$\alpha_3=\tfrac{1}{2}$ and $\alpha_4=1$ are the stage coefficients.
\section{Numerical results}
\label{sec:test-cases}
In order to validate the meshless HLLC approximate Riemann solver proposed in
this paper, Toro's one-dimensional gas shock tube problem is chosen as the first
benchmark test. The method's capability to deal with the compressible water in a
liquid shock tube is also inspected (Except this test, all the other cases are
compressible gas flow problems).  Then more complicated two-dimensional problems
including moving shock passing a cylinder, internal supersonic flows in channels
and external transonic flows over aerofoils are examined. 
All the following numerical simulations are performed on a Compaq
Presario-CQ45 laptop equipped with a dual-core P4 2.16GHz CPU, 1GB memory and
250GB hard drive. The main operating system on the laptop is Windows Vista
(32bit), but the computation is run under Cygwin, which provides a Linux like
environment.  The underlying serial computing programs are written in Fortran 90
language and compiled by the GNU Fortran compiler. In this work, the finite
volume method is employed together with the JST scheme, which is based on a
central differencing with second order and fourth order artificial dissipation
\cite{1981-Jameson-p1259}, to produce the reference results.

\subsection{Gas shock tube}
Toro's shock tube, of which the mathematical model is the one-dimensional Euler
equations, is chosen as the first test case. Although it is only a 1D problem,
the exact solution is quite complicated as it consists of a shock wave, a
contact discontinuity and a rarefaction. It provides a critical examination of
the numerical method's capability to resolve these complex physical
features. The tube is filled with ideal gases of different states separated by
an imaginary barrier imposed at the centre initially. The density, velocity and
pressure of the gases at the initial state are given by
\begin{equation}
  \mathbf{W}=
  \left\{
  \begin{array}{lr}
    (1, 0.75, 1) & -0.5\le x \le 0 \\
    (0.125, 0, 0.1) & 0< x \le 0.5
  \end{array}
  \right.
\end{equation}

\begin{figure}[ht]
  \centering
  \includegraphics[width=0.6\textwidth]{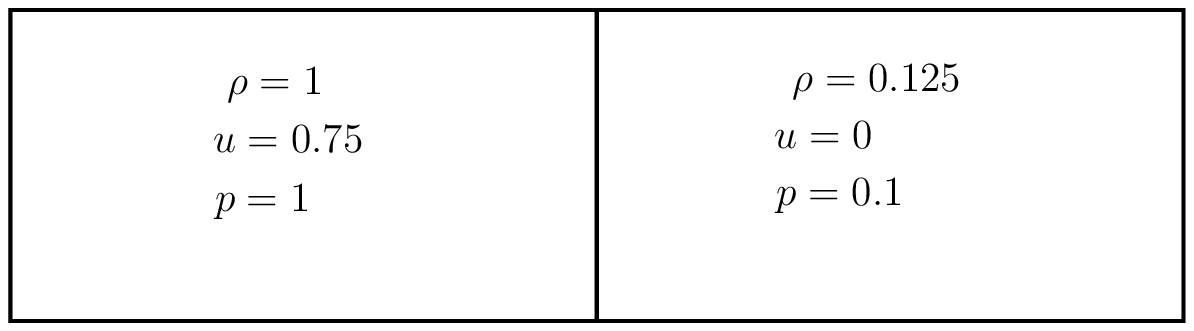}
  \caption{Initial condition setup for the gas shock tube}
  \label{fig:gas:shock:init}
\end{figure}
\begin{figure}[ht]
  \centering
  \subfigure[density]{\includegraphics[width=0.48\textwidth]{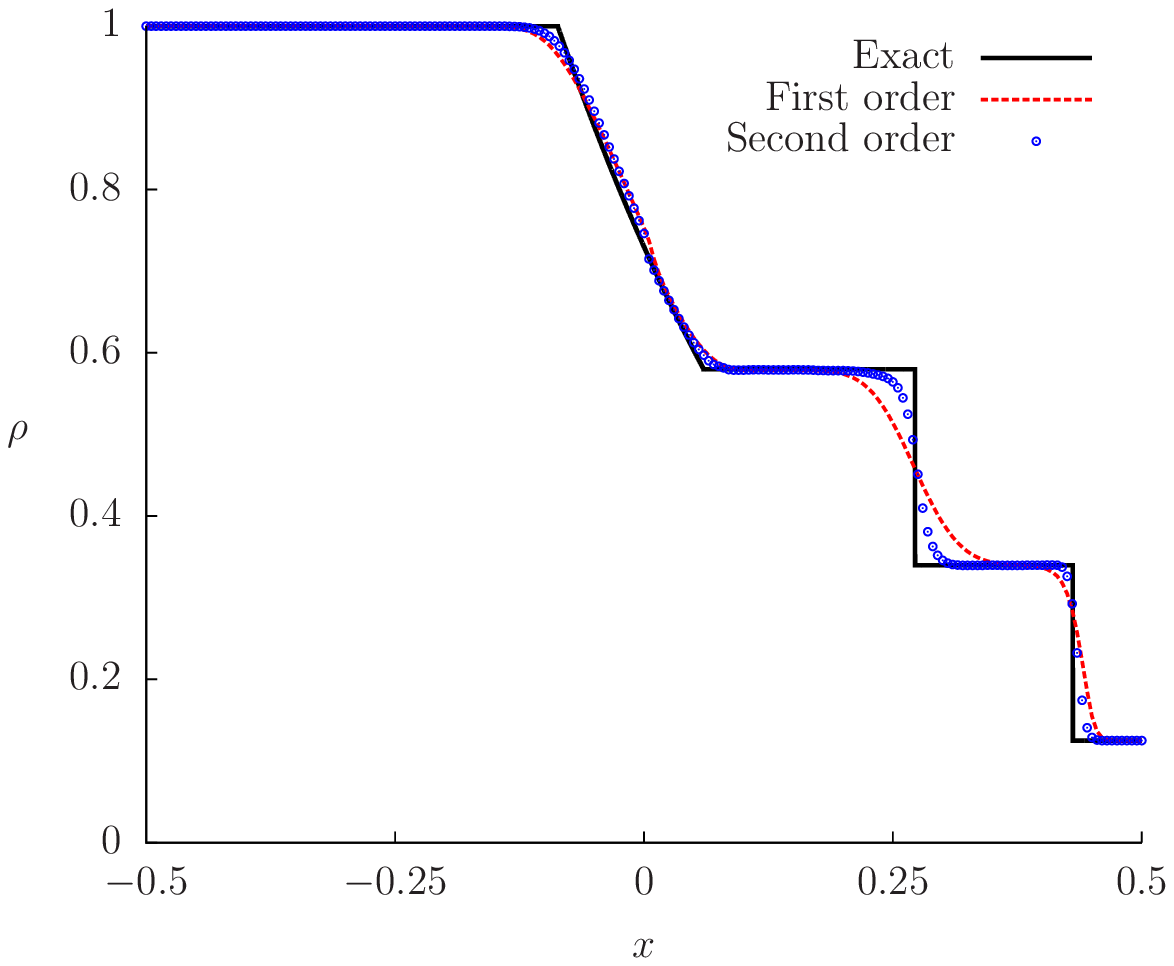}}
  \subfigure[velocity]{\includegraphics[width=0.48\textwidth]{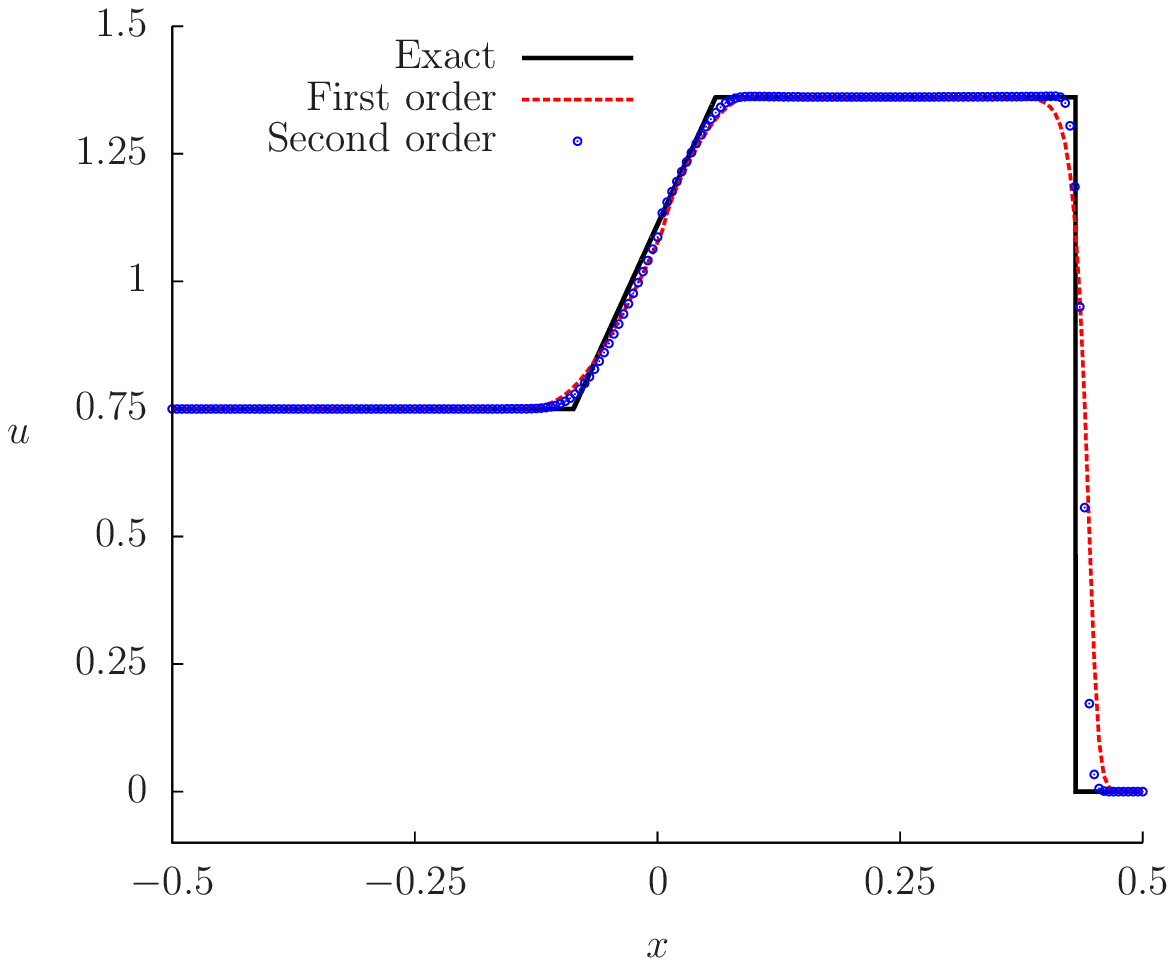}}
  \subfigure[pressure]{\includegraphics[width=0.48\textwidth]{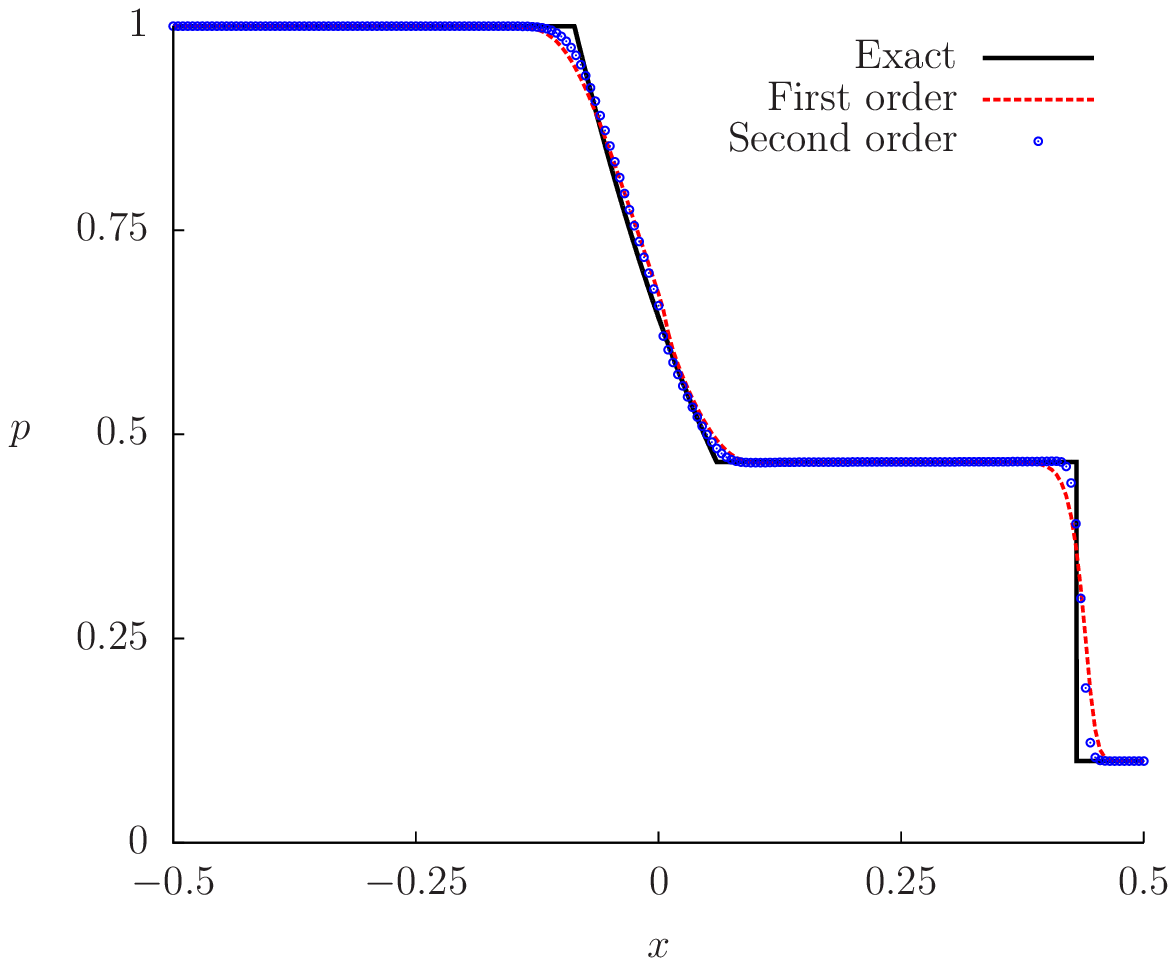}}
  \subfigure[internal energy]{\includegraphics[width=0.48\textwidth]{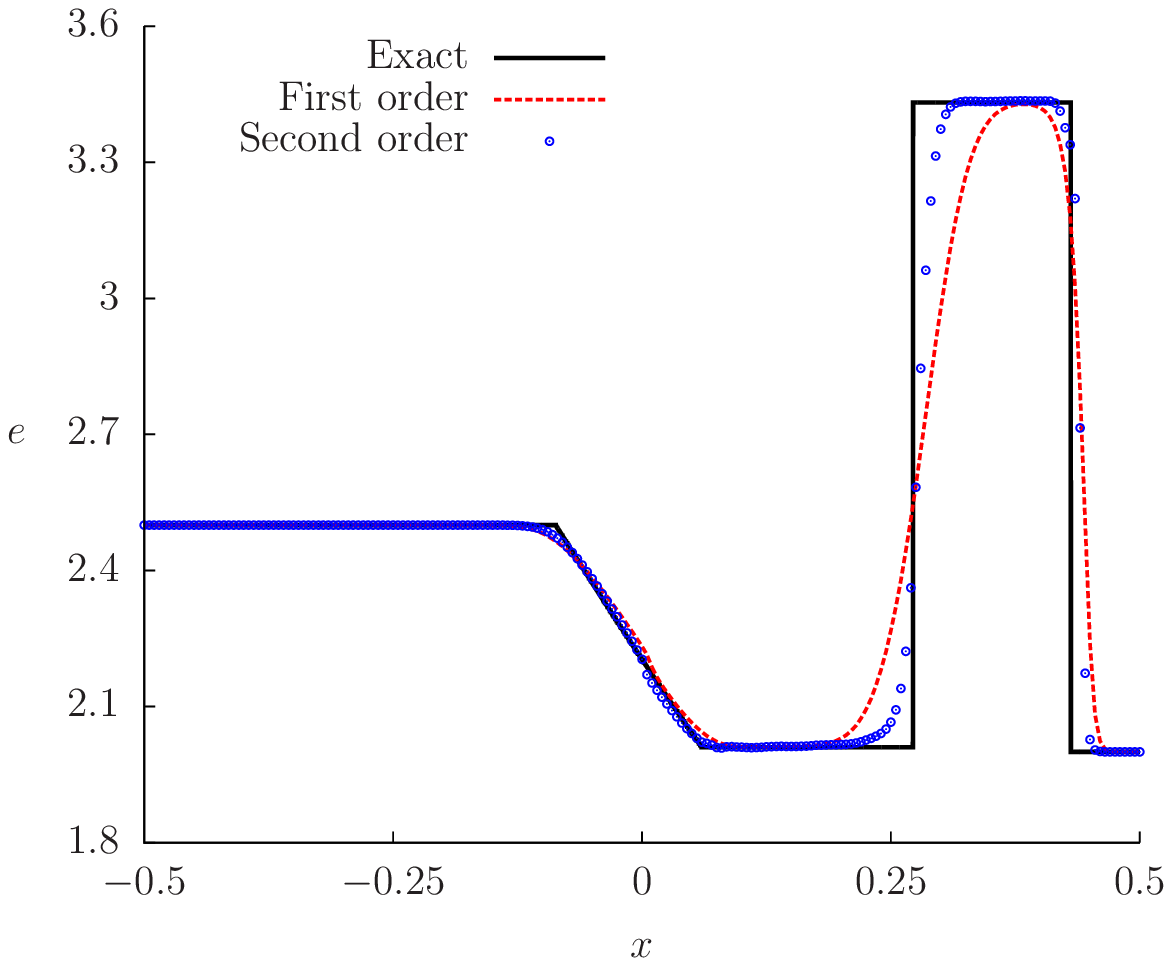}}
  \caption{Gas shock tube}
  \label{fig:shock:tube:gas:1}
\end{figure}

Transmissive boundary conditions are applied at $x=-0.5$ and $0.5$.  Four
hundred meshless points are used to divide the shock tube uniformly and the
solution is computed to $t=0.2$. Both the first order Godunov scheme and
piecewise linear reconstruction strategy are used to obtain the numerical
results. The exact solution is produced by the computer program listed in
Chapter 4 of Toro's book \cite{Book-1997-Toro}. The program utilises an
iterative guess-correction (Newton-Raphson) strategy to find the exact solution,
while ten thousands mesh points are used to depict the exact solution in the
present work. The initial condition setup is illustrated in Figure
\ref{fig:gas:shock:init}. The density, velocity, pressure and density-averaged
internal energy of the numerical and exact solutions are plotted in Figure
\ref{fig:shock:tube:gas:1}, in which the solid line represents the exact
solution, the dashed line is the first order result and the circular symbol
stands for second order result. It clearly shows that the contact discontinuity
(at about $x=0.25$) is successfully captured by the meshless HLLC approximate
Riemann solver. The second-order scheme provides a reasonable better resolution
of the contact discontinuity than the first-order one. The shock wave (at about
$x=0.4$) is also resolved well by the numerical method.

\subsection{Liquid shock tube}
The second test case is a liquid shock tube problem proposed by Ivings et
al. \cite{Paper-1998-Ivings-395}. The density, velocity and pressure of the
liquids at the initial states are given by
\begin{equation}
  \mathbf{W}=
  \left\{
  \begin{array}{lr}
    (1100, 500, 5\times 10^9) & -0.5\le x \le 0 \\
    (1000, 0, 10^5) & 0< x \le 0.5
  \end{array}
  \right.
\end{equation}

\begin{figure}[ht]
  \centering
  \includegraphics[width=0.6\textwidth]{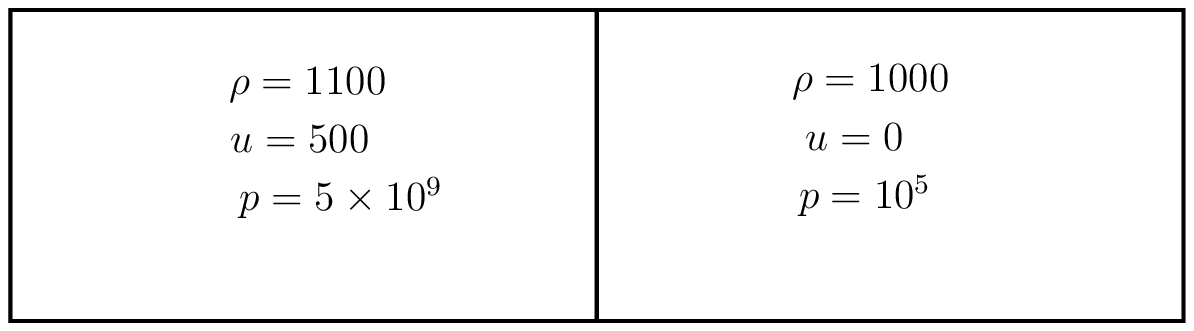}
  \caption{Initial condition setup for the liquid shock tube}
  \label{fig:liquid:shock:init}
\end{figure}
\begin{figure}[ht]
  \centering
  \subfigure[density]{\includegraphics[width=0.48\textwidth]{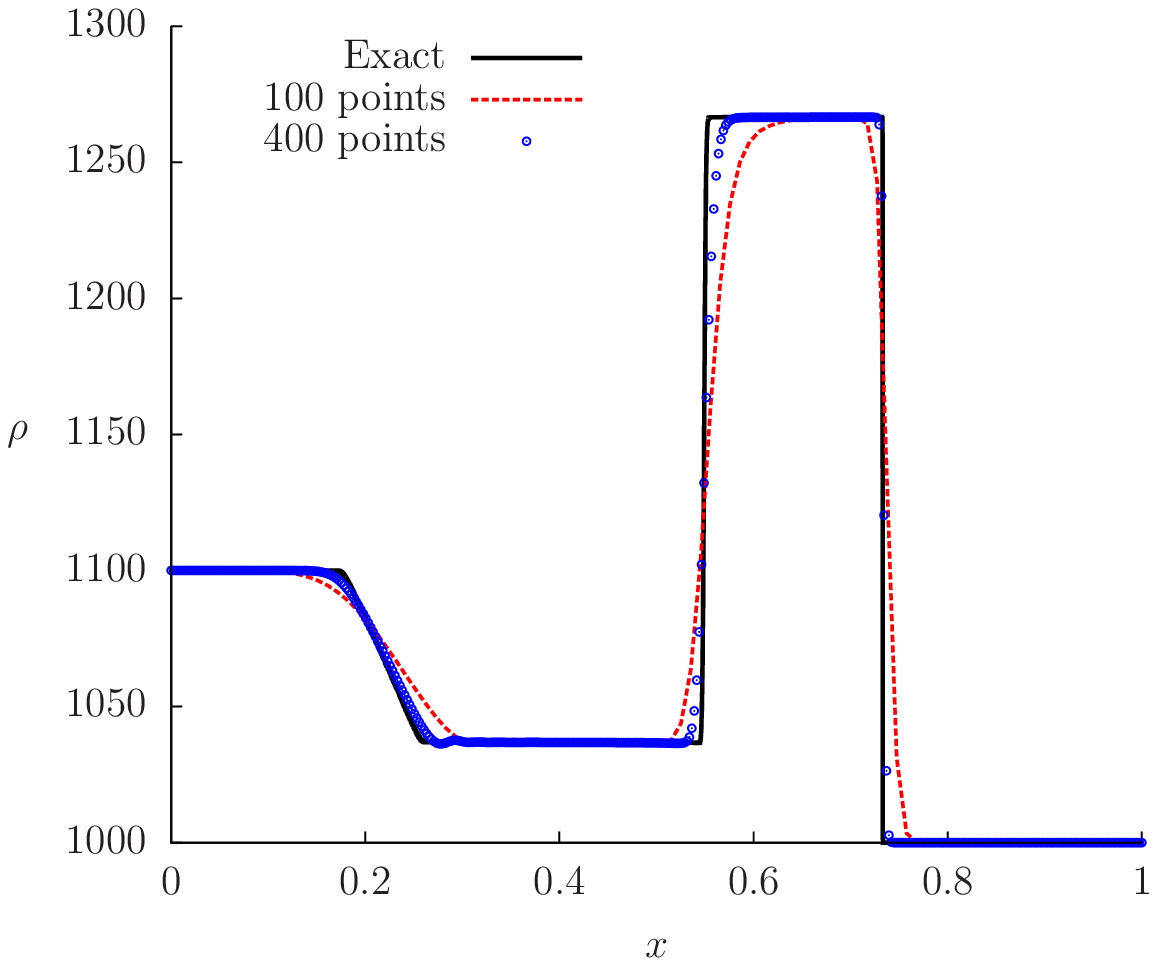}}
  \subfigure[velocity]{\includegraphics[width=0.48\textwidth]{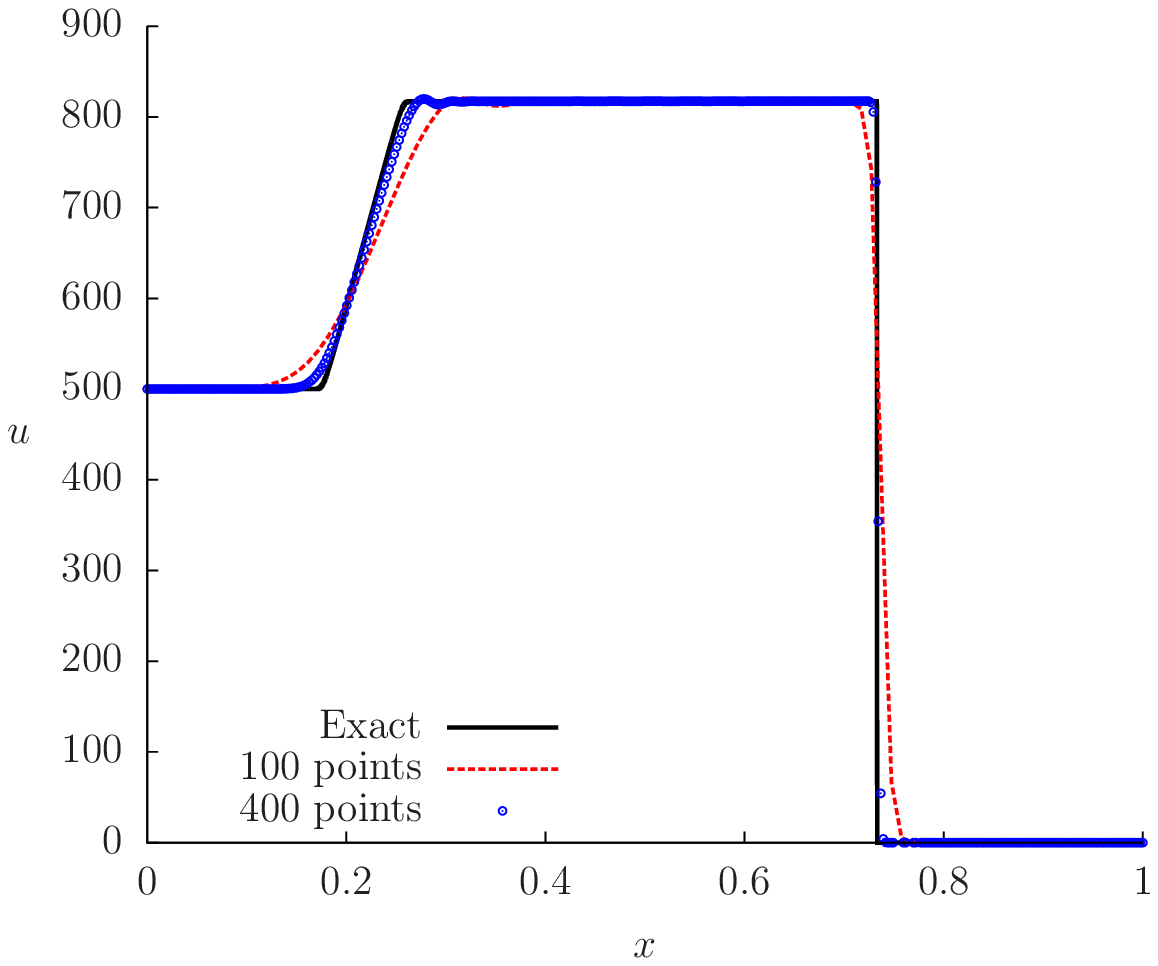}}
  \subfigure[pressure]{\includegraphics[width=0.48\textwidth]{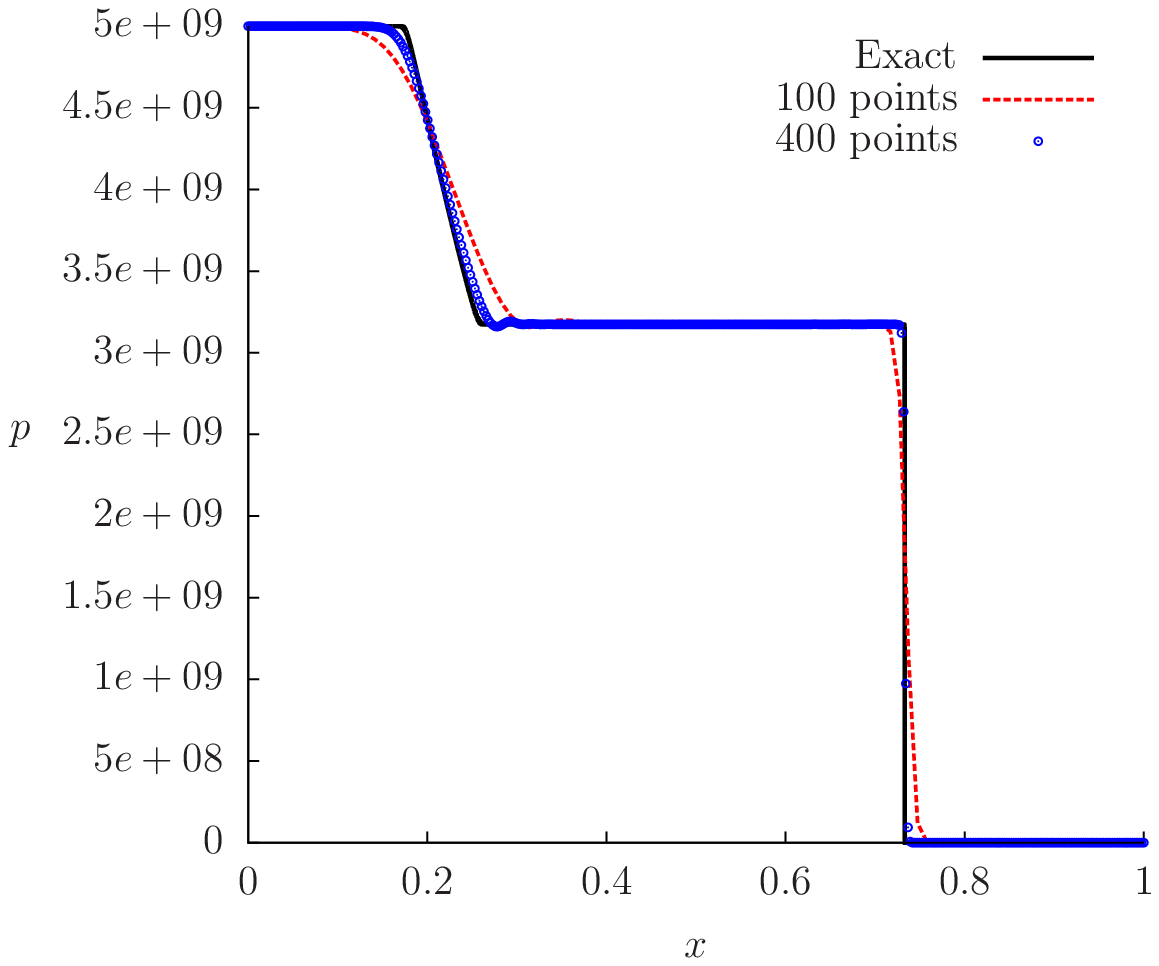}}
  \subfigure[internal energy]{\includegraphics[width=0.48\textwidth]{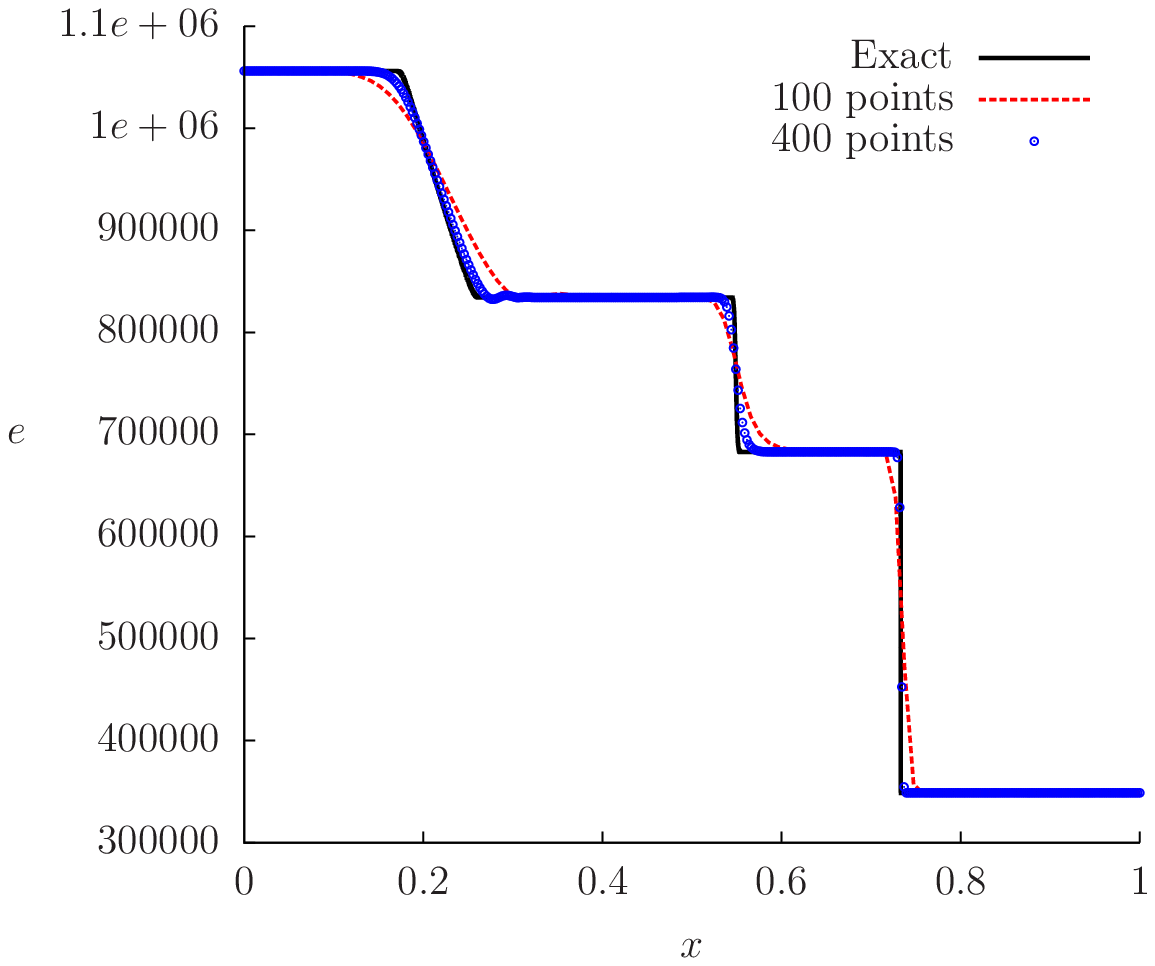}}
  \caption{Liquid shock tube}
  \label{fig:shock:tube:liquid:1}
\end{figure}

The polytropic constant is set to $\gamma=7.15$ and the pressure constant is
taken as $p_{\rm c}=3\times 10^8$. The flow field is advanced to $t=6\times
10^{-5}$. The initial condition setup is shown in Figure
\ref{fig:liquid:shock:init}. Figure \ref{fig:shock:tube:liquid:1} shows the
exact solution and the numerical solutions. The exact solution is depicted by
ten thousand points, the numerical solutions are obtained by the second order
scheme for 100 and 400 meshless particles respectively. Obviously, the numerical
method provides satisfactory results considering the wave speeds, locations and
strengths.

After testing the numerical method for one-dimensional problems, we continue to
benchmark the method for two-dimensional problems. 

\subsection{Moving shock passing a cylinder}
\begin{figure}[ht]
  \centering
  \includegraphics[width=0.6\textwidth]{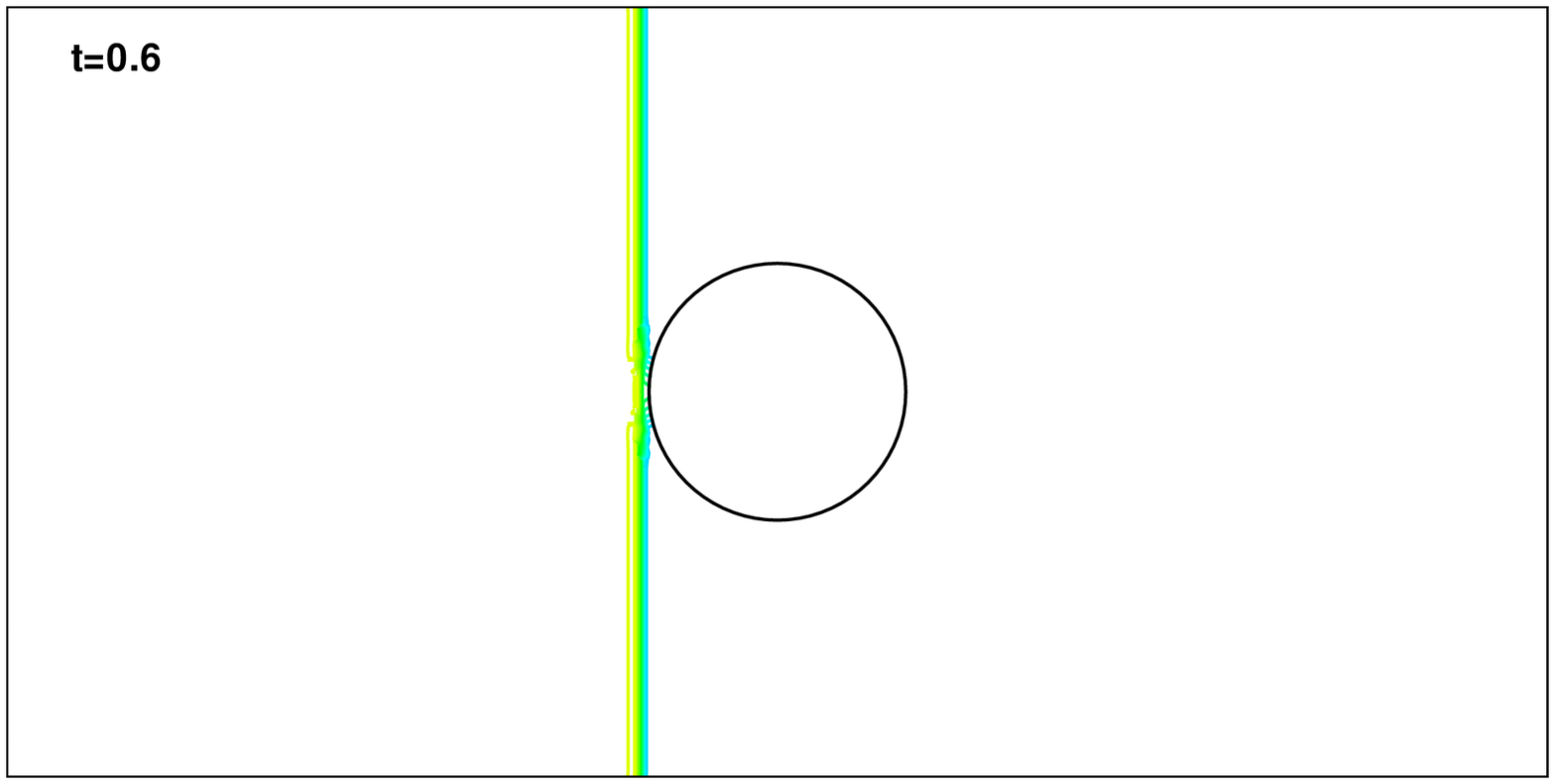}
  \includegraphics[width=0.6\textwidth]{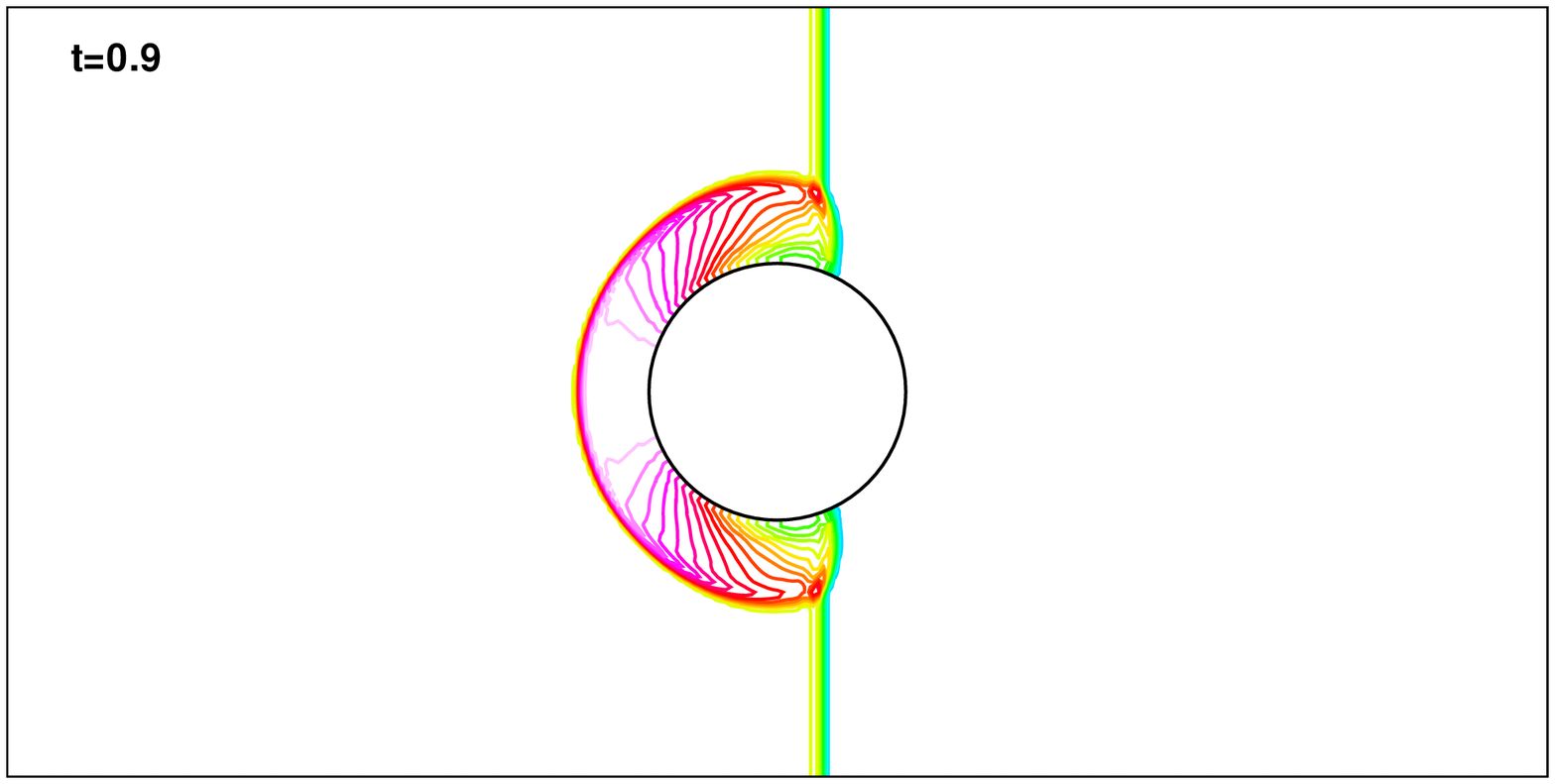}
  \includegraphics[width=0.6\textwidth]{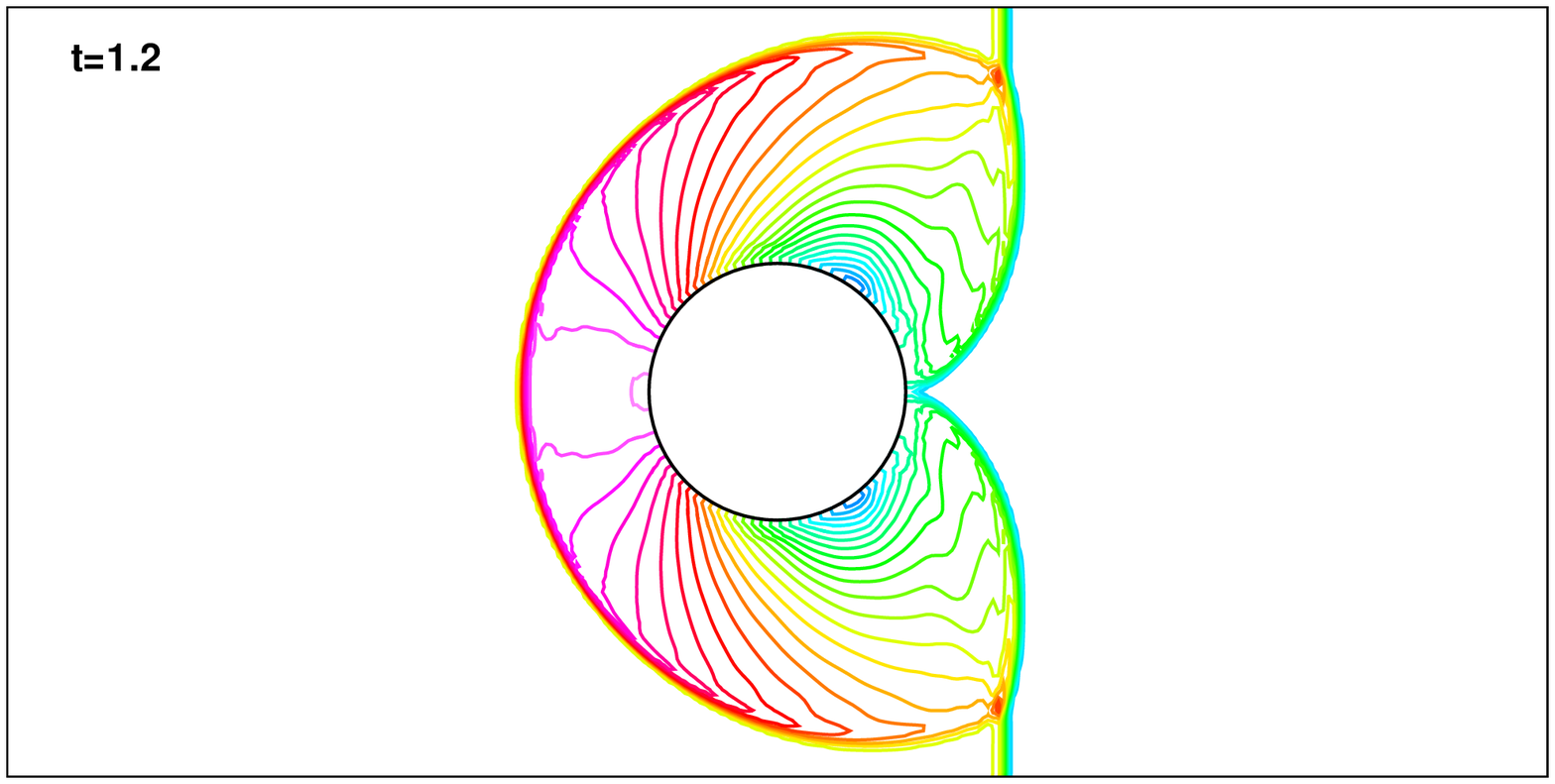}
  \includegraphics[width=0.6\textwidth]{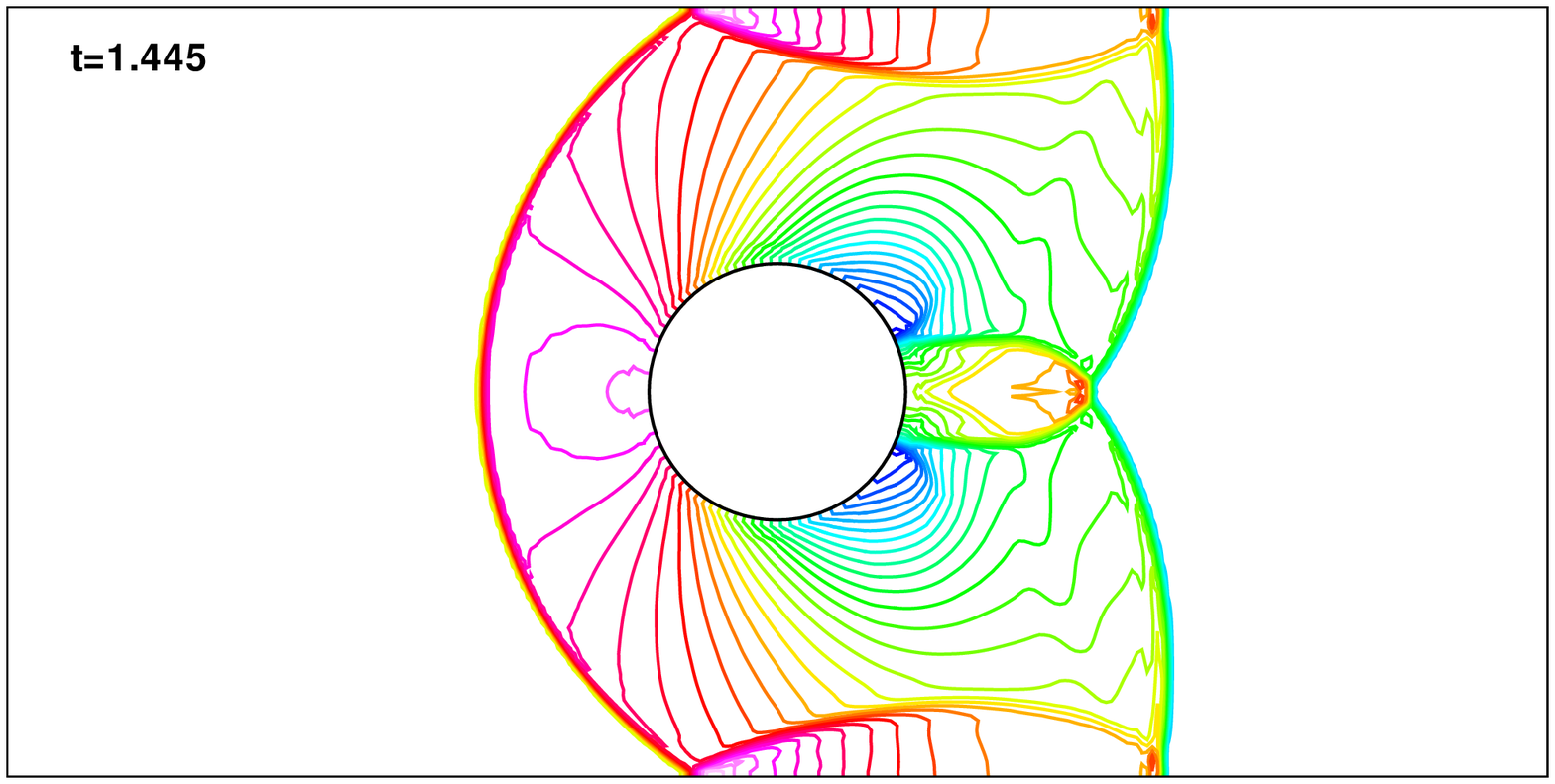}
  \caption{Moving shock wave passing a cylinder}
\label{fig:shock:structure}
\end{figure}

This case was proposed by Luo et al. \cite{2006-Luo-p618}, in which an incident
shock with $M_s=2$ is moving from the left side to the right side and passing a
cylinder in a rectangular tube. The length of the tube is $6$ and its height is
3. The cylinder of radius $r=0.5$ is located at the centre of the tube.  The
number of the points distributed in the domain is $43,438$.  The solution is
advanced to $t=1.445$. Figure \ref{fig:shock:structure} exhibits the snapshots
of the flow field at $t=0.6$, $0.9$, $1.2$ and $1.445$ with the density
contours. Once the shock hits the cylinder ($t=0.6$), it is reflected to the
left, upward and downward directions ($t=0.9\sim 1.2$). When the upward and
downward reflected shocks touch the top and bottom boundaries of the tube, they
are reflected again($t=1.445$). These complicated flow features are clearly
shown in Figure \ref{fig:shock:structure}. The present solution agrees well with
the result of Luo et al. (Figure 19 of \cite{2006-Luo-p618}).

\subsection{Internal supersonic flow}
\begin{figure}[ht]
  \centering
  \subfigure[Mach number contours - FVM]{\includegraphics[width=0.6\textwidth]{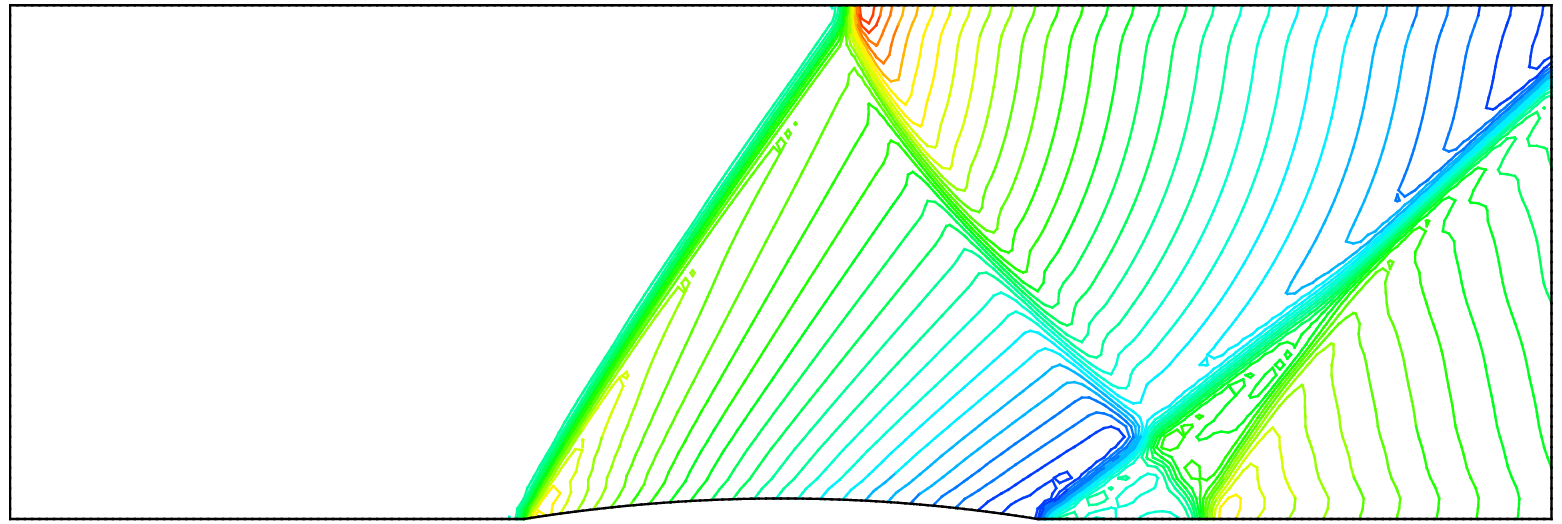}}
  \subfigure[Mach number contours - LSMM]{\includegraphics[width=0.6\textwidth]{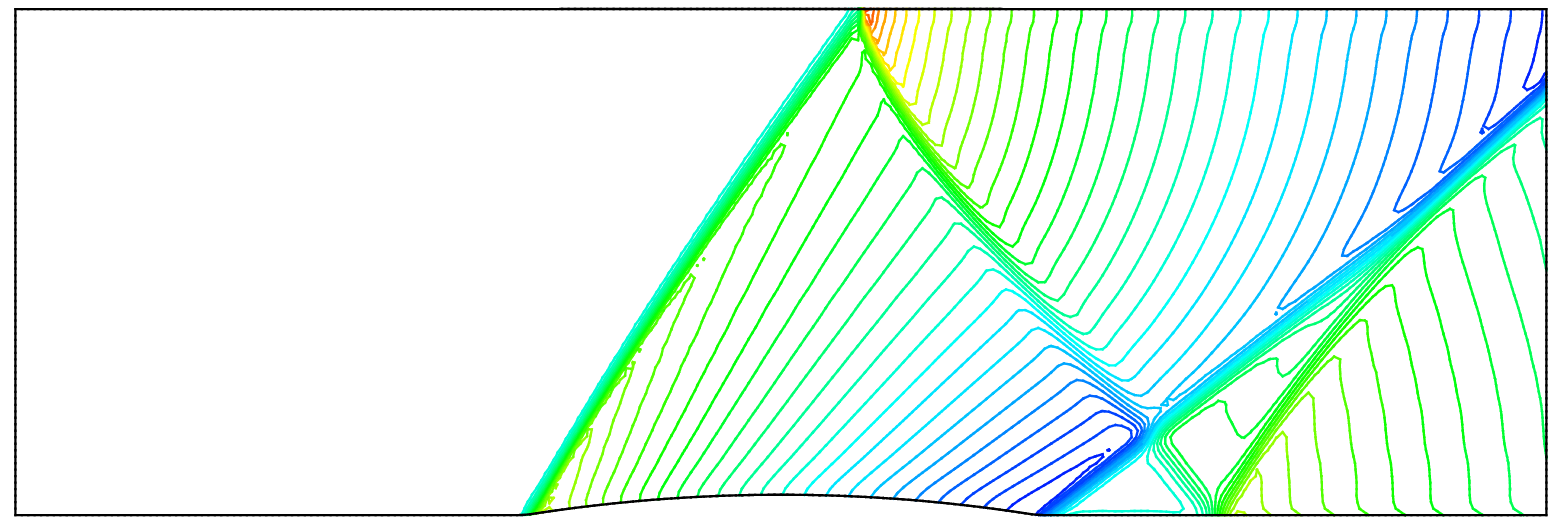}}
  \subfigure[Pressure coefficient]{\includegraphics[width=0.6\textwidth]{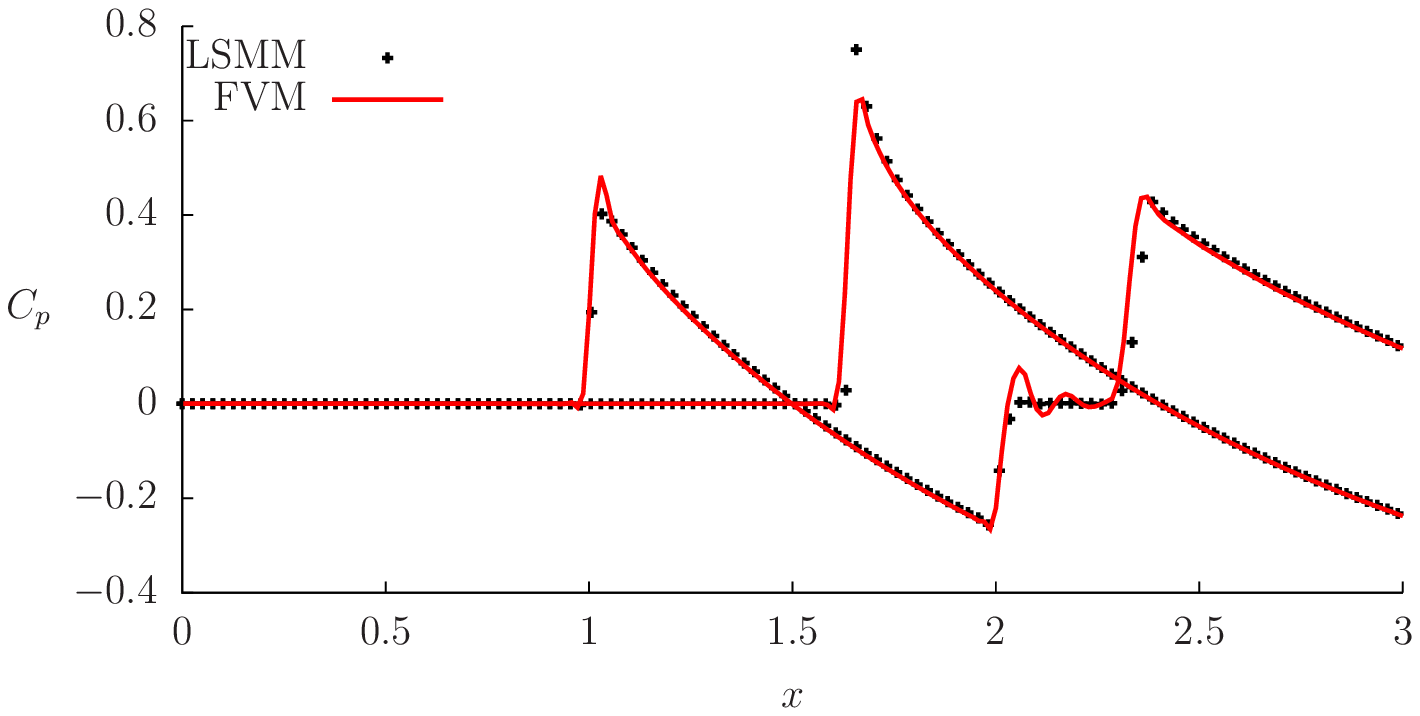}}
  \caption{Supersonic flow in a tube with a $4\%$ circular bump, $M_{\infty}=1.4$}
  \label{fig:supersonic:flow:bump}
\end{figure}

Supersonic gas flow of $M_{\infty}=1.4$ in a channel with a $4\%$ circular bump
is considered here. The length of the channel is $3$ and its height is $1$, the
circular bump is located at the centre of the bottom boundary of the channel.
The number of the points in the domain is $19,517$.  The solution is started
with the uniform supersonic flow and then advanced by the four-stage explicit
Runge-Kutta method to the steady state. Besides the meshless method, we also
compute the solution with the finite volume method for the purpose of
comparison.  The results of FVM and LSMM are shown in Figure
\ref{fig:supersonic:flow:bump}. Inspecting the Mach number contours, it is not
difficult to find that the flow features captured by LSMM are identical to those
captured by FVM. The FVM solution exhibits strong oscillations near the shocks
while LSMM result avoids these as shown in Figure \ref{fig:supersonic:flow:bump}
(c), which depicts the pressure coefficients on the top and bottom boundaries of
the channel.

\subsection{Transonic flows over the NACA0012 aerofoil}
\label{sec:transonic-flows-over}

Two groups of transonic flows over the symmetric NACA0012 aerofoil
with different flow conditions are chosen. The flow conditions for the first set
are $M_{\infty}=0.8$ and $\alpha=0^{\rm o}$. For the second set, they are
$M_{\infty}=0.8$ and $\alpha=1.25^{\rm o}$.  

\begin{figure}[ht]
  \centering
  \includegraphics[width=0.48\textwidth]{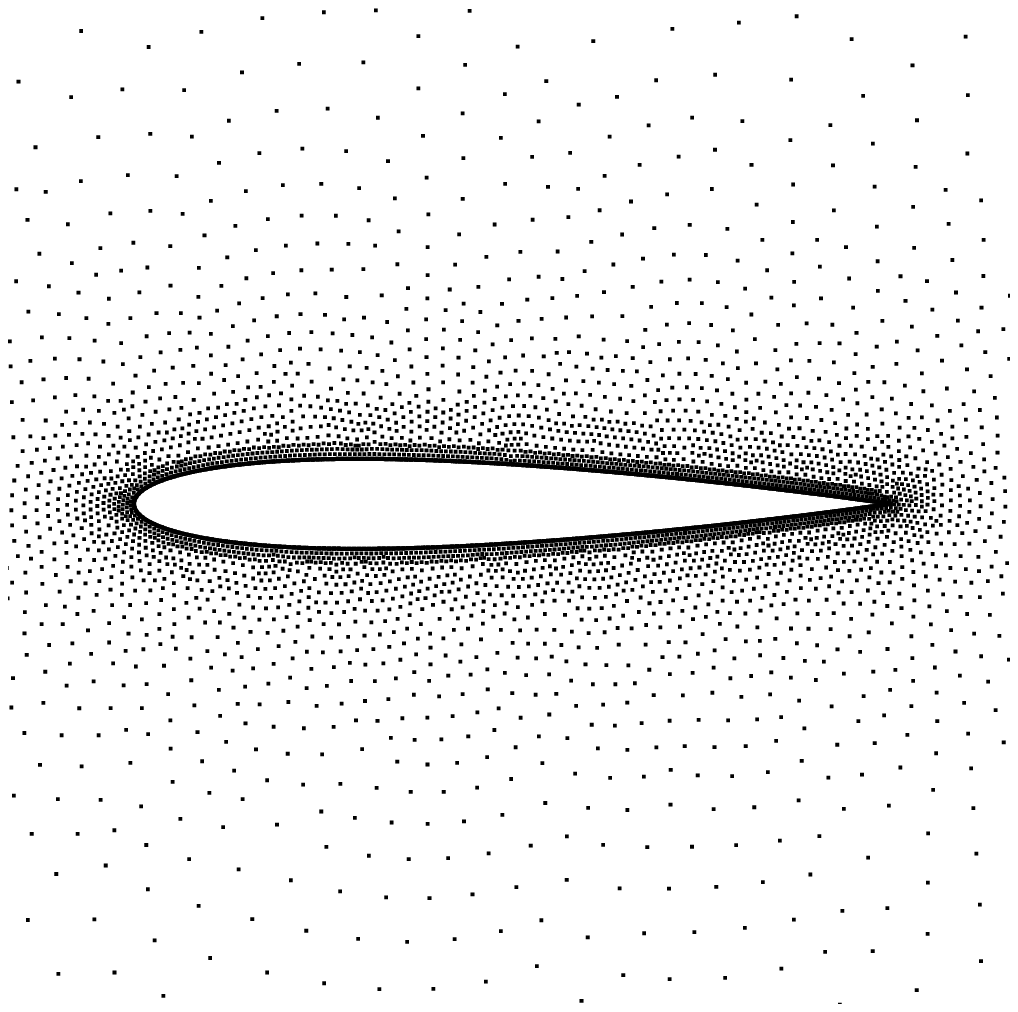}
  \caption{Meshless points distribution around the NACA0012 aerofoil}
  \label{fig:naca0012:point}
\end{figure}

The number of meshless points in the flow domain is 5,557, while there are 42 points on
the far field boundary and 310 nodes on the wall boundary. The natural
neighbours around every point are chosen to form the meshless clouds. Due to the
random property of the point distribution, the number of satellites for each
cloud are not the same.  For this case, the maximum number of satellites is 9,
the minimum value is 4 and more than ninety percent of the clouds have six or seven
satellites.  Figure \ref{fig:naca0012:point} exhibits the scattered points around
the NACA0012 aerofoil.

The physical domain is initialised with uniform flows with the corresponding
Mach number and angle of attack. Then the numerical solution starts with the
impulse posed by the aerofoil. Local time stepping is applied to enhance the
convergence rate to obtain the steady flows. 

Figure \ref{fig:naca0012:m080:contour} shows the Mach number contours in
the flow field for these two-group tests. The first row presents the results of
the zero-angle-of-attack and the second row give the results for $\alpha=1.25^{\rm
  o}$. We organise the FVM solution in the left column and the LSMM solution in
the right column. For the case of zero-angle-of-attack, two strong shock waves
symmetric about the aerofoil chord ($x$-axis) appear in the flow field, the
locations of the shock wave are at the half chord. For
$\alpha=1.25^{\rm o}$, a strong shock appears in the upper part of the domain and a
relatively weak shock appears in the lower part of the field. The LSMM solution
and FVM solution are almost the same as displayed in
\ref{fig:naca0012:m080:contour}. 

\begin{figure}[ht]
  \centering
  \subfigure[$M_{\infty}=0.8$, $\alpha=0^{\rm o}$]{
     \includegraphics[width=0.48\textwidth]{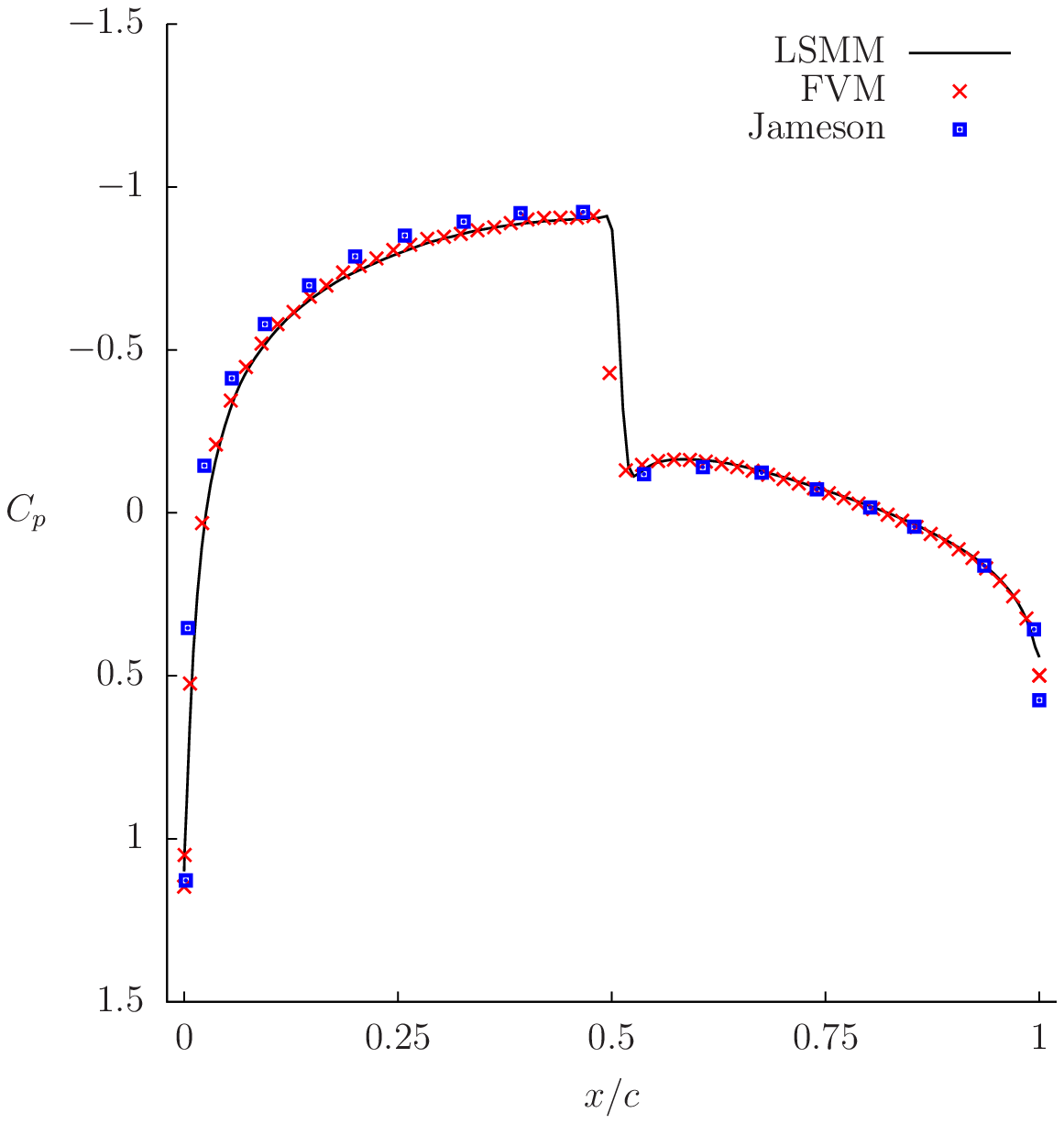}}
  \subfigure[$M_{\infty}=0.8$, $\alpha=1.25^{\rm o}$]{
     \includegraphics[width=0.48\textwidth]{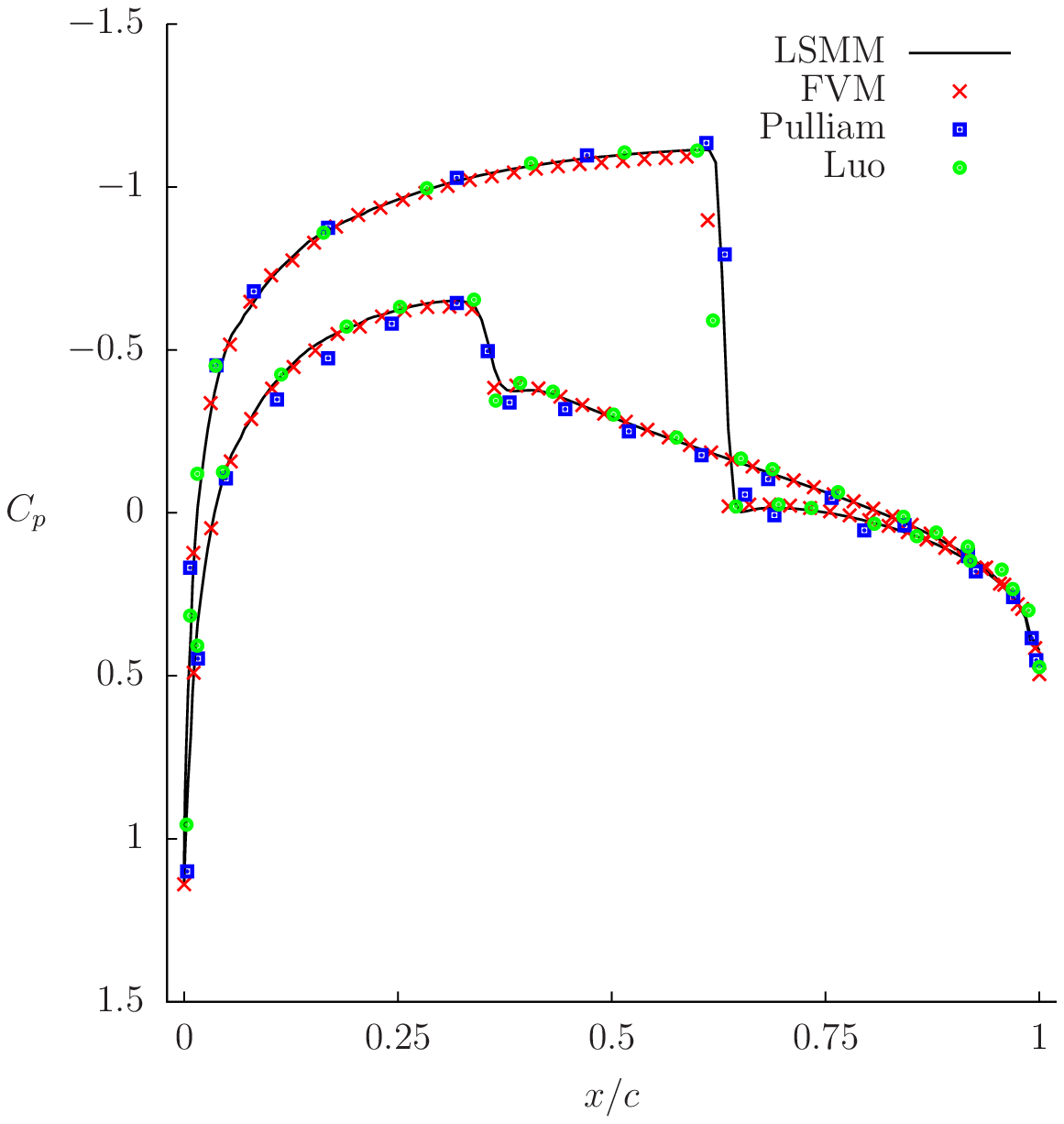}}
  \caption{Pressure coefficient around the NACA0012 aerofoil}
  \label{fig:naca0012:m080:Cp}
\end{figure}

\begin{figure}[ht]
  \centering
  \subfigure[$M_{\infty}=0.8$, $\alpha=0^{\rm o}$]{
     \includegraphics[width=0.48\textwidth]{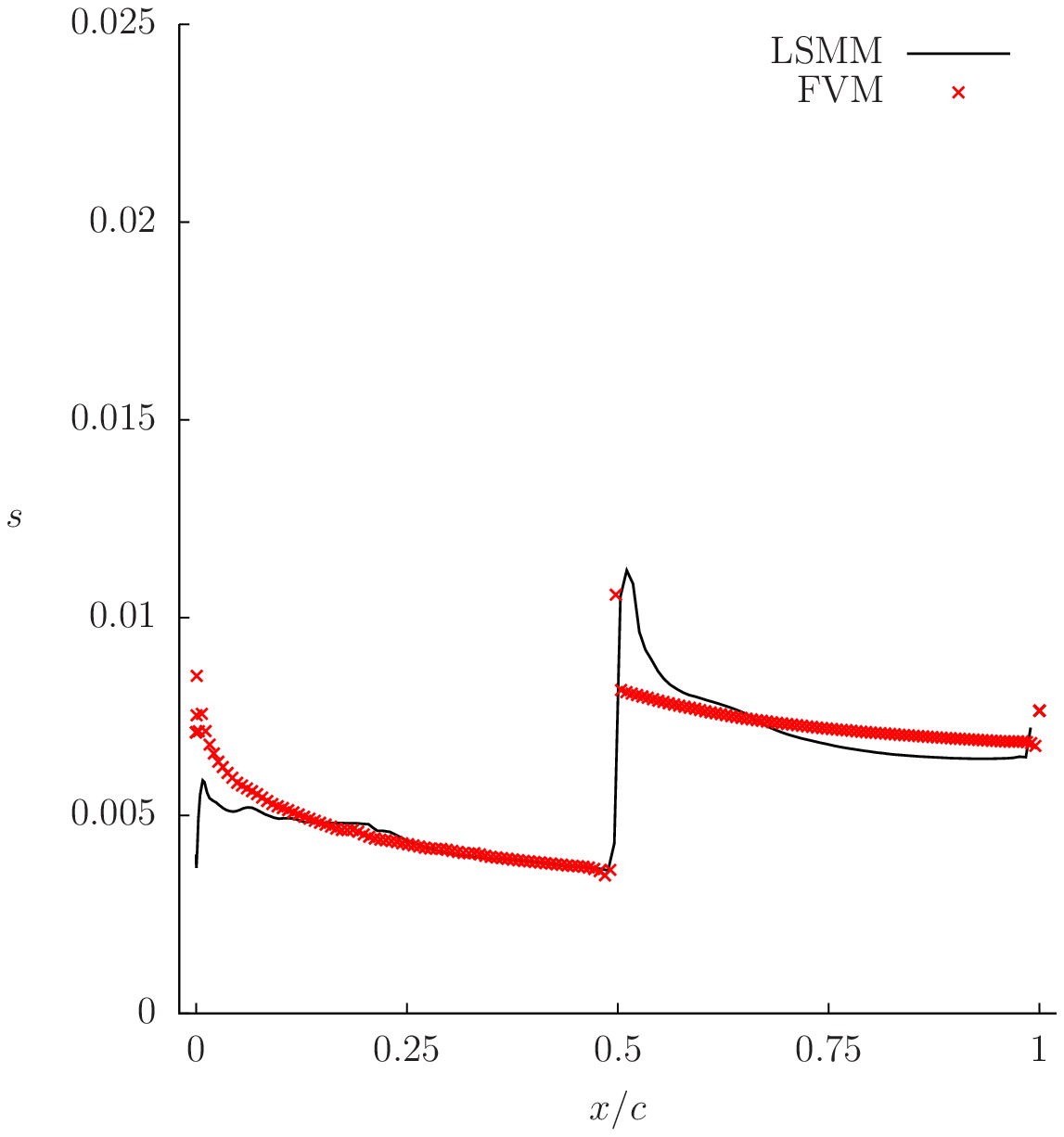}}
  \subfigure[$M_{\infty}=0.8$, $\alpha=1.25^{\rm o}$]{
     \includegraphics[width=0.48\textwidth]{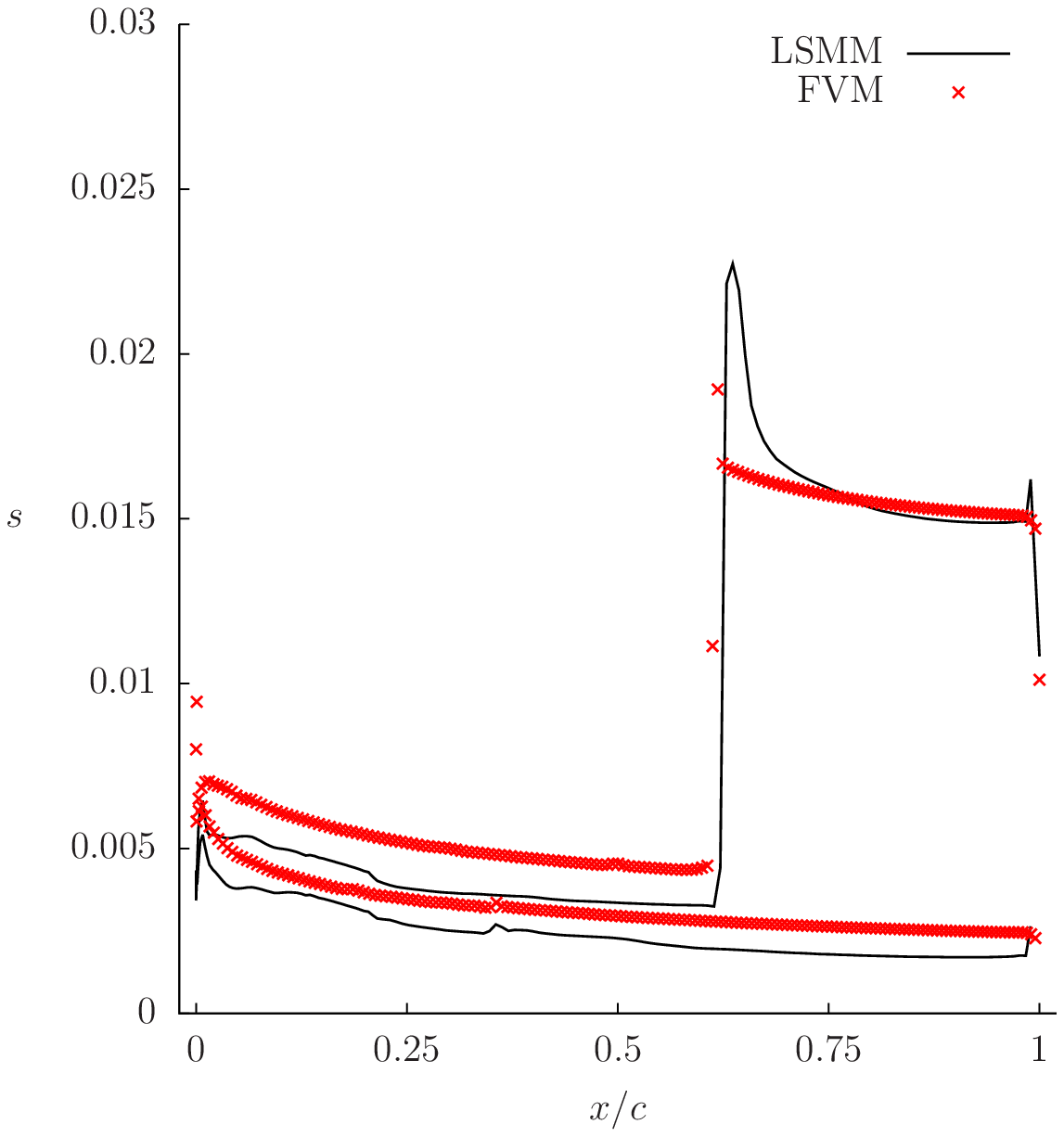}}
  \caption{Entropy production around the NACA0012 aerofoil}
  \label{fig:naca0012:m080:s}
\end{figure}

\begin{figure}[htp]
  \centering
  \subfigure[$M_{\infty}=0.8$, $\alpha=0^{\rm o}$ - FVM]{
     \includegraphics[width=0.48\textwidth]{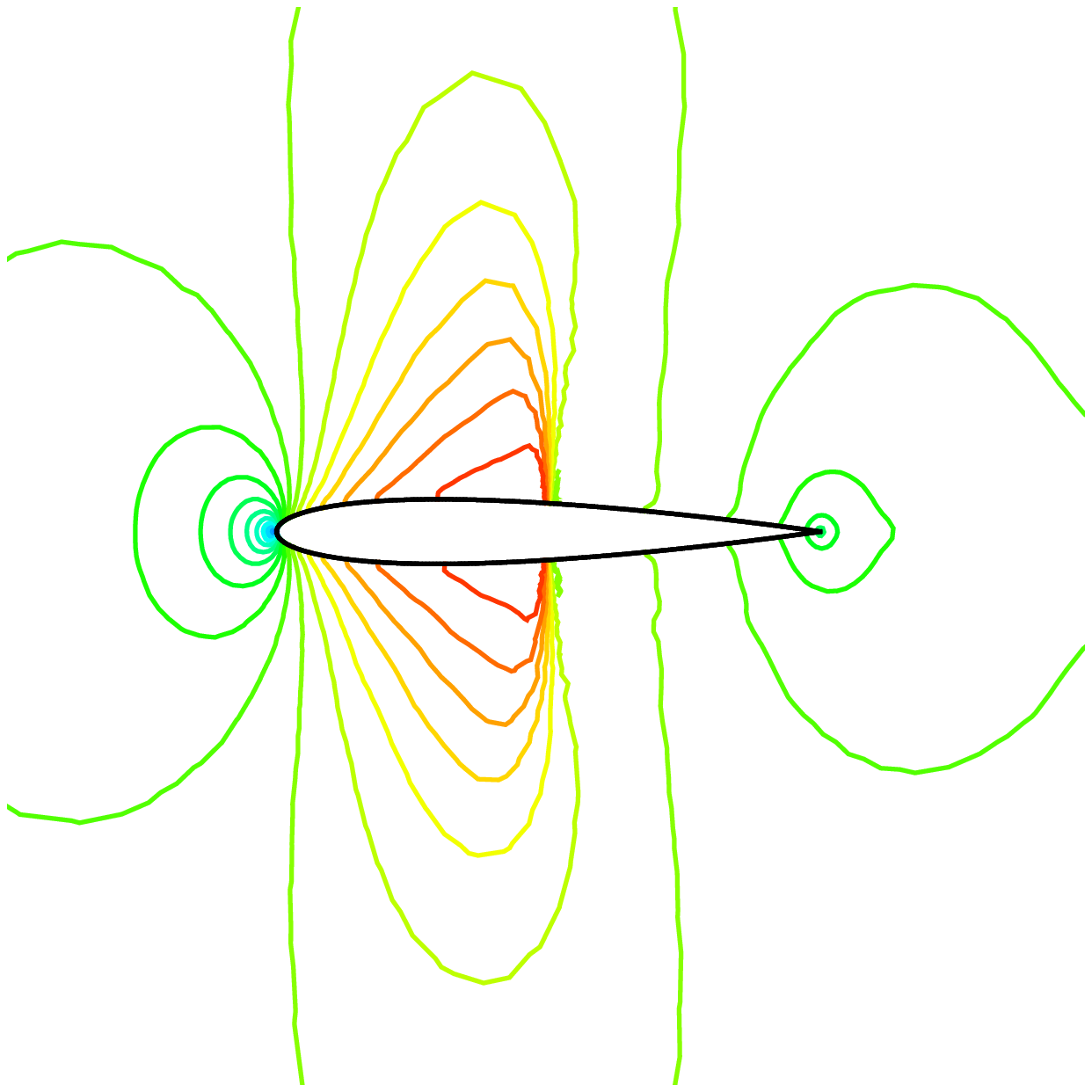}}
  \subfigure[$M_{\infty}=0.8$, $\alpha=0^{\rm o}$ - LSMM]{
     \includegraphics[width=0.48\textwidth]{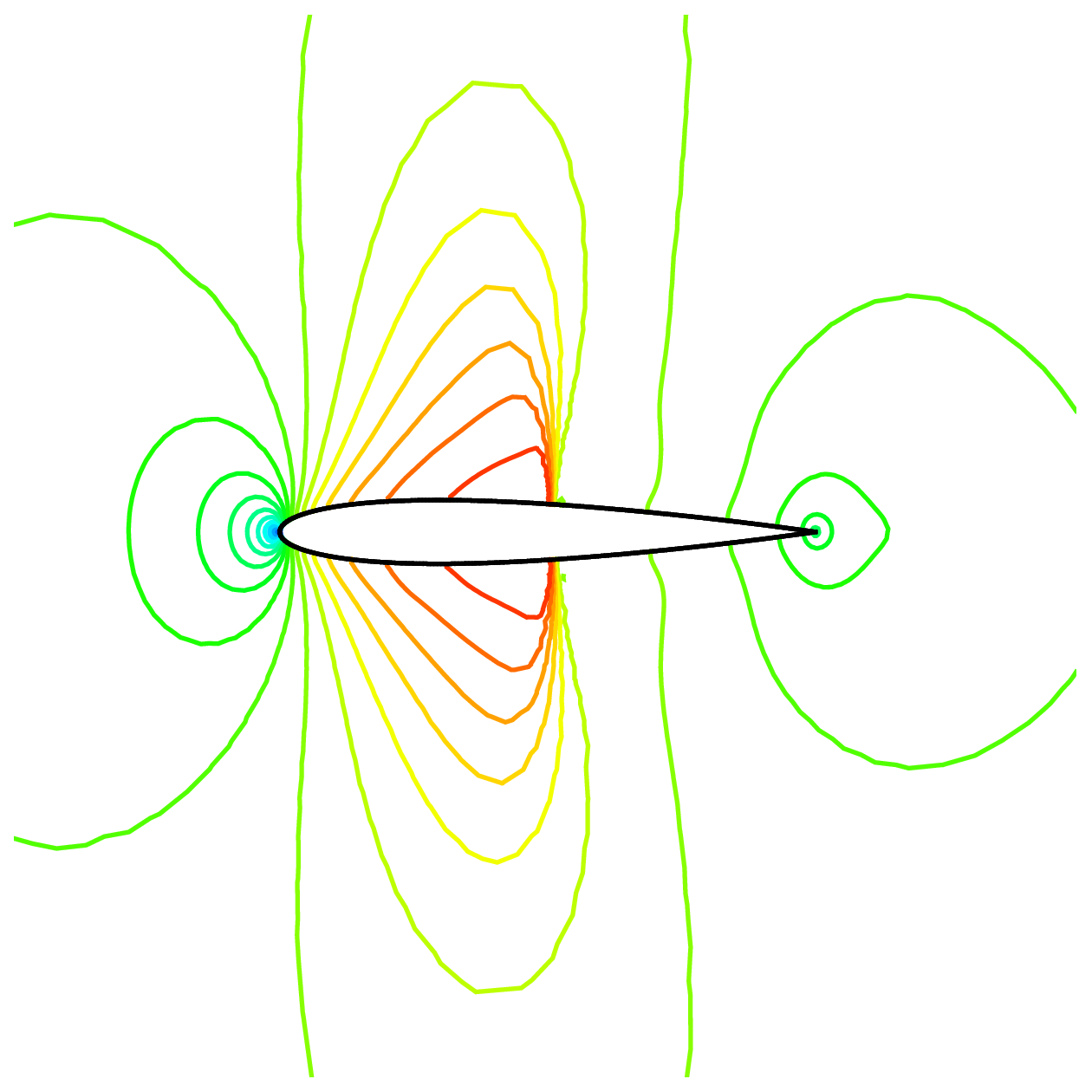}}
  \subfigure[$M_{\infty}=0.8$, $\alpha=1.25^{\rm o}$ - FVM]{
     \includegraphics[width=0.48\textwidth]{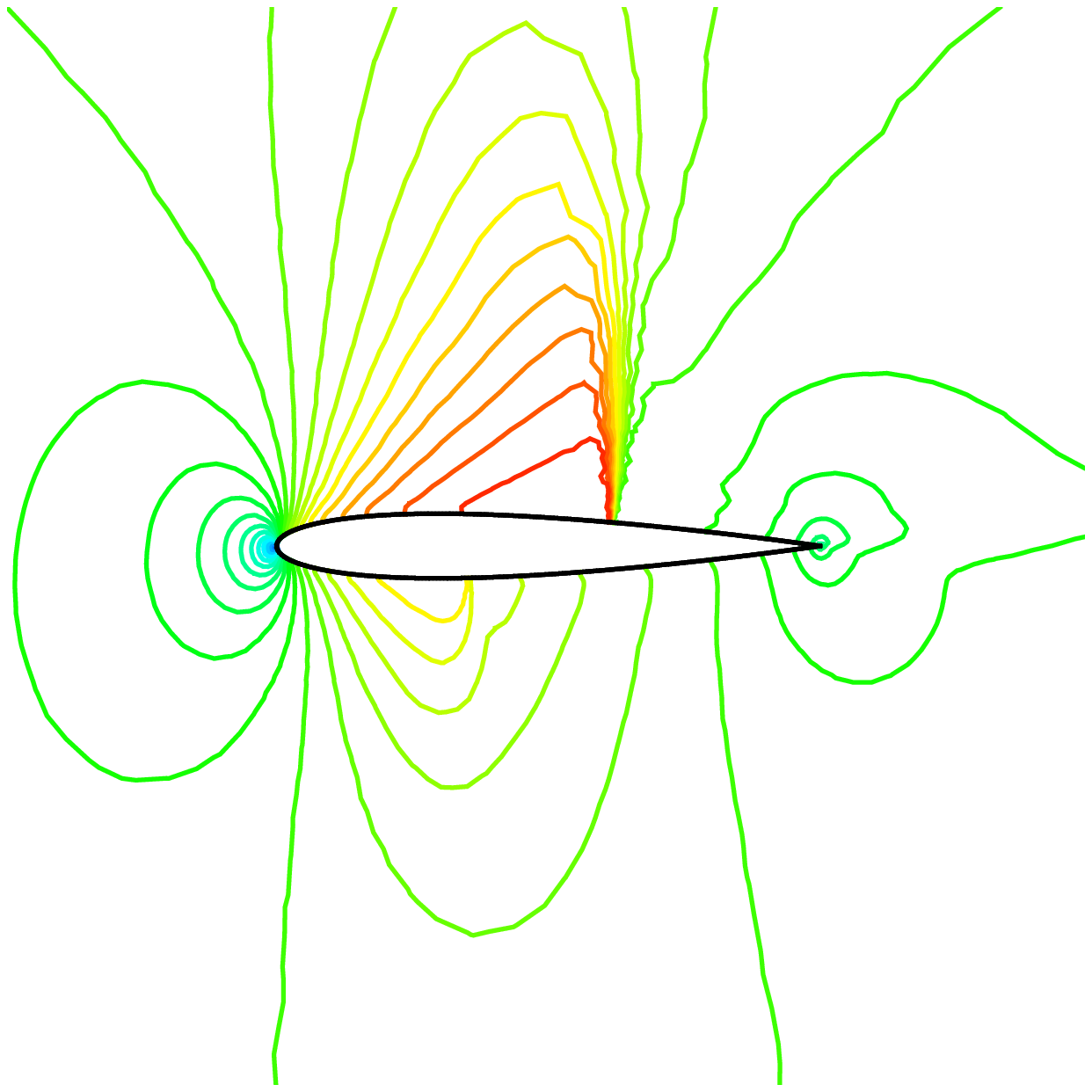}}
  \subfigure[$M_{\infty}=0.8$, $\alpha=1.25^{\rm o}$ - LSMM]{
     \includegraphics[width=0.48\textwidth]{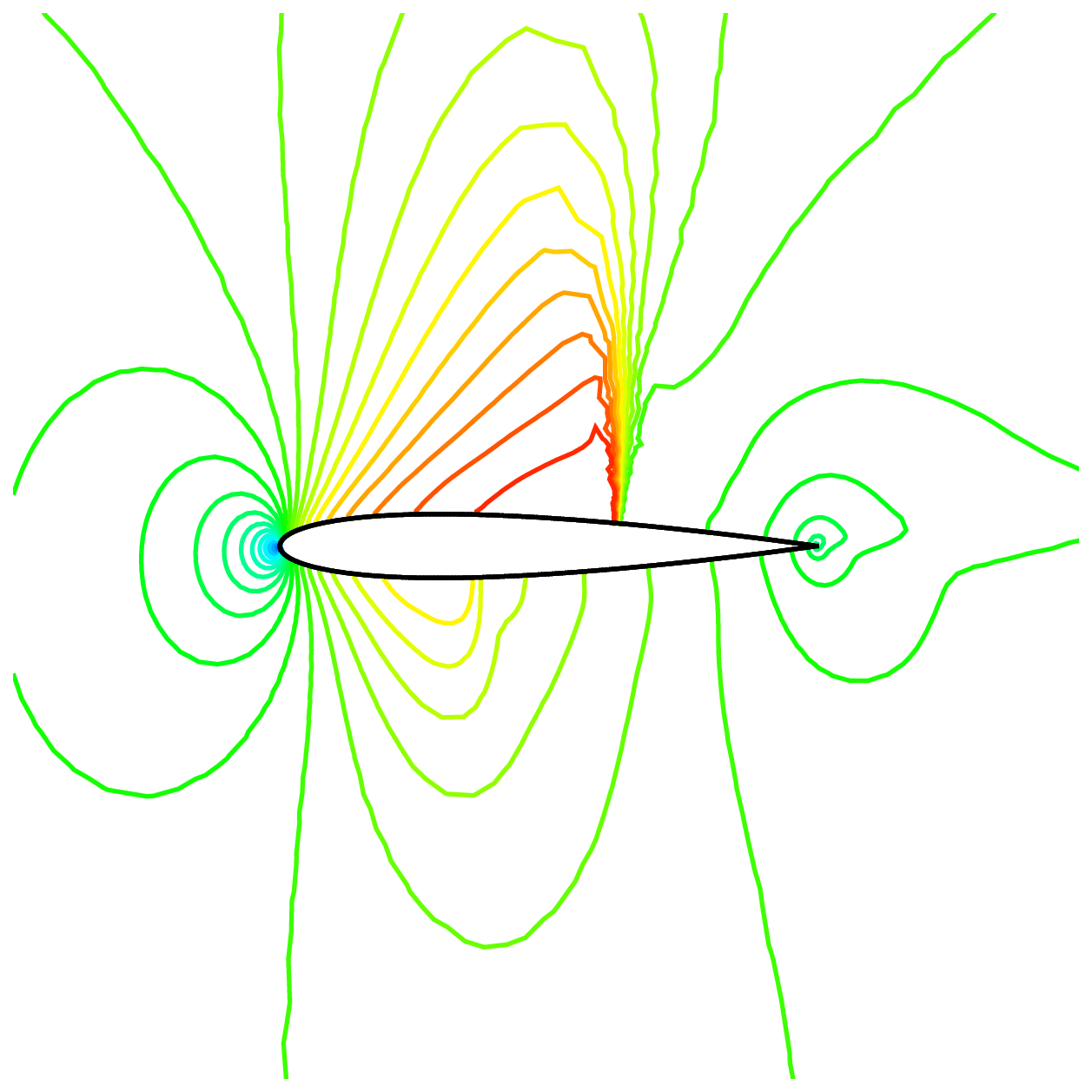}}
  \caption{Mach number contours for the NACA0012 aerofoil}
  \label{fig:naca0012:m080:contour}
\end{figure}

More details can be found when we look at Figure \ref{fig:naca0012:m080:Cp}, in
which the pressure coefficients around the aerofoil are depicted.  The LSMM
solutions are in good agreement with the FVM and other reference results
(Jameson et al. \cite{1981-Jameson-p1259}, Pulliam and Stegeret
\cite{Conference-1985-Pulliam} and Luo et al \cite{Paper-2008-Luo-597}) not only
in smooth regions but also in regions with large gradients. The entropy productions 
on the aerofoil by FVM and LSMM are shown in Fig. \ref{fig:naca0012:m080:s}. For these
two cases, it seems that LSMM performs slightly better at the leading edge 
as it gives smaller entropy productions than FVM does.

\subsection{Transonic flow over the RAE2822 aerofoil}
\label{sec:transonic-RAE2822}
\begin{figure}[H]
  \centering
  \includegraphics[width=0.48\textwidth]{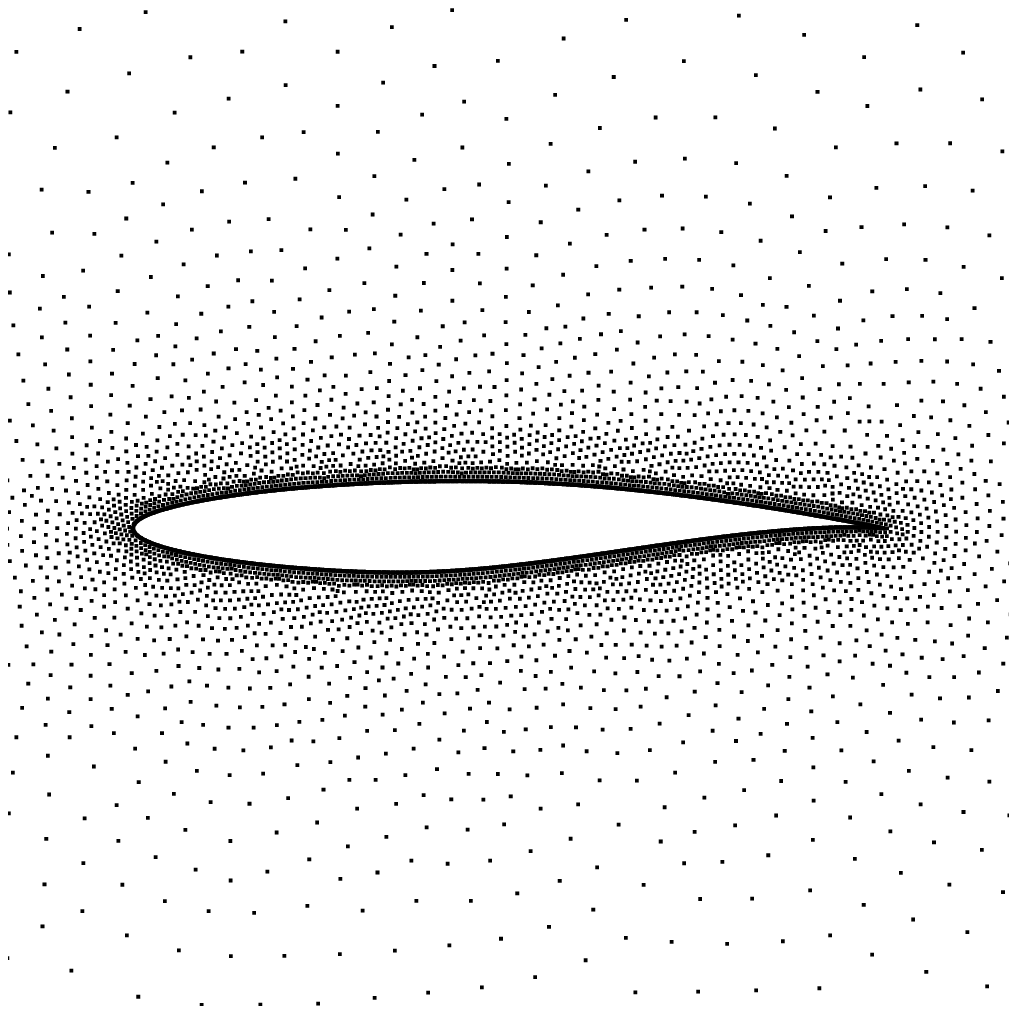}
  \caption{Meshless points distribution around the RAE2822 aerofoil}
  \label{fig:rae2822:point}
\end{figure}

\begin{figure}[H]
  \centering
  \subfigure[Pressure coefficient]{    \label{fig:rae2822:cp}
    \includegraphics[width=0.48\textwidth]{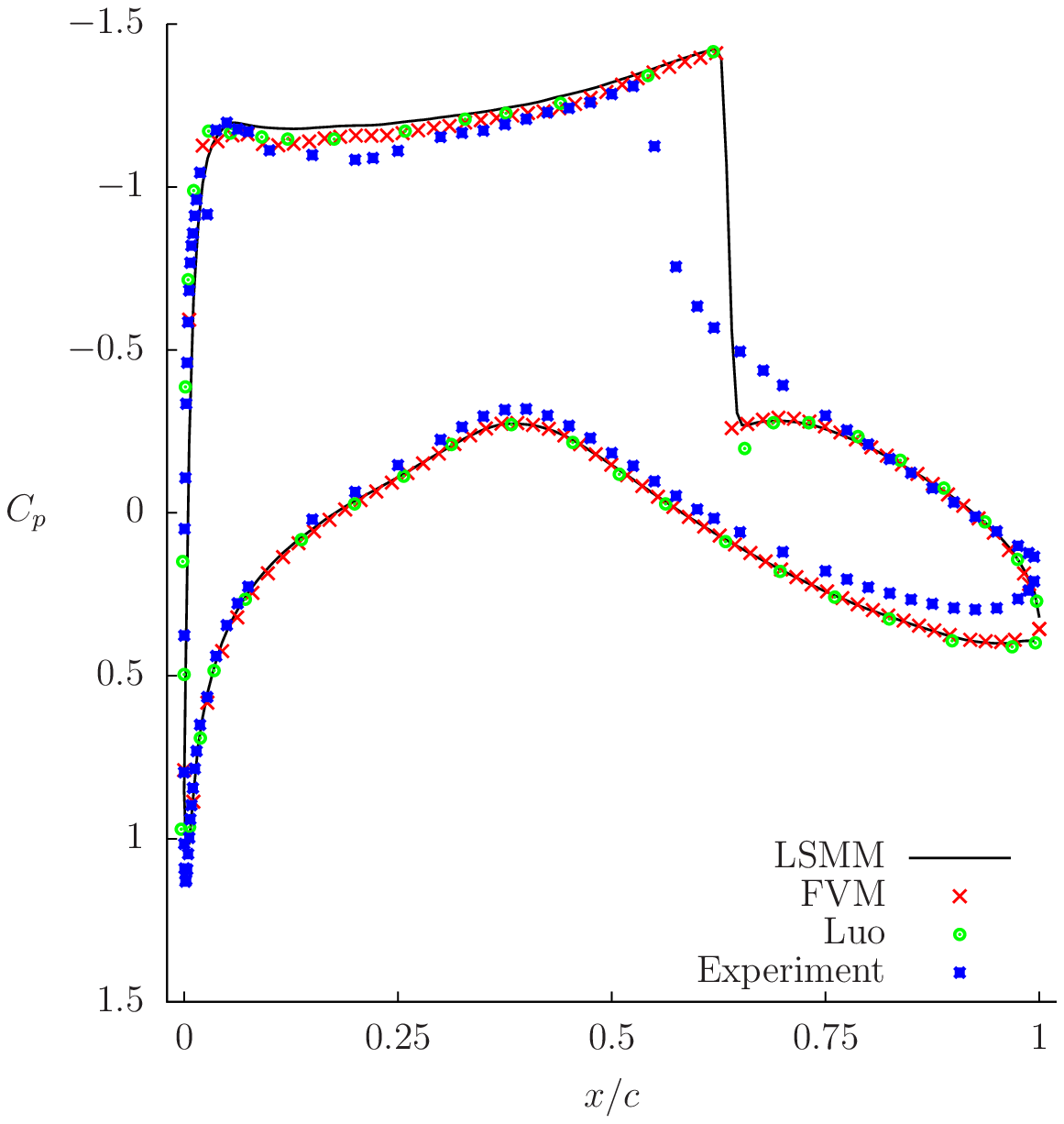}}\quad
  \subfigure[Entropy production]{\label{fig:rae2822:s}  
    \includegraphics[width=0.48\textwidth]{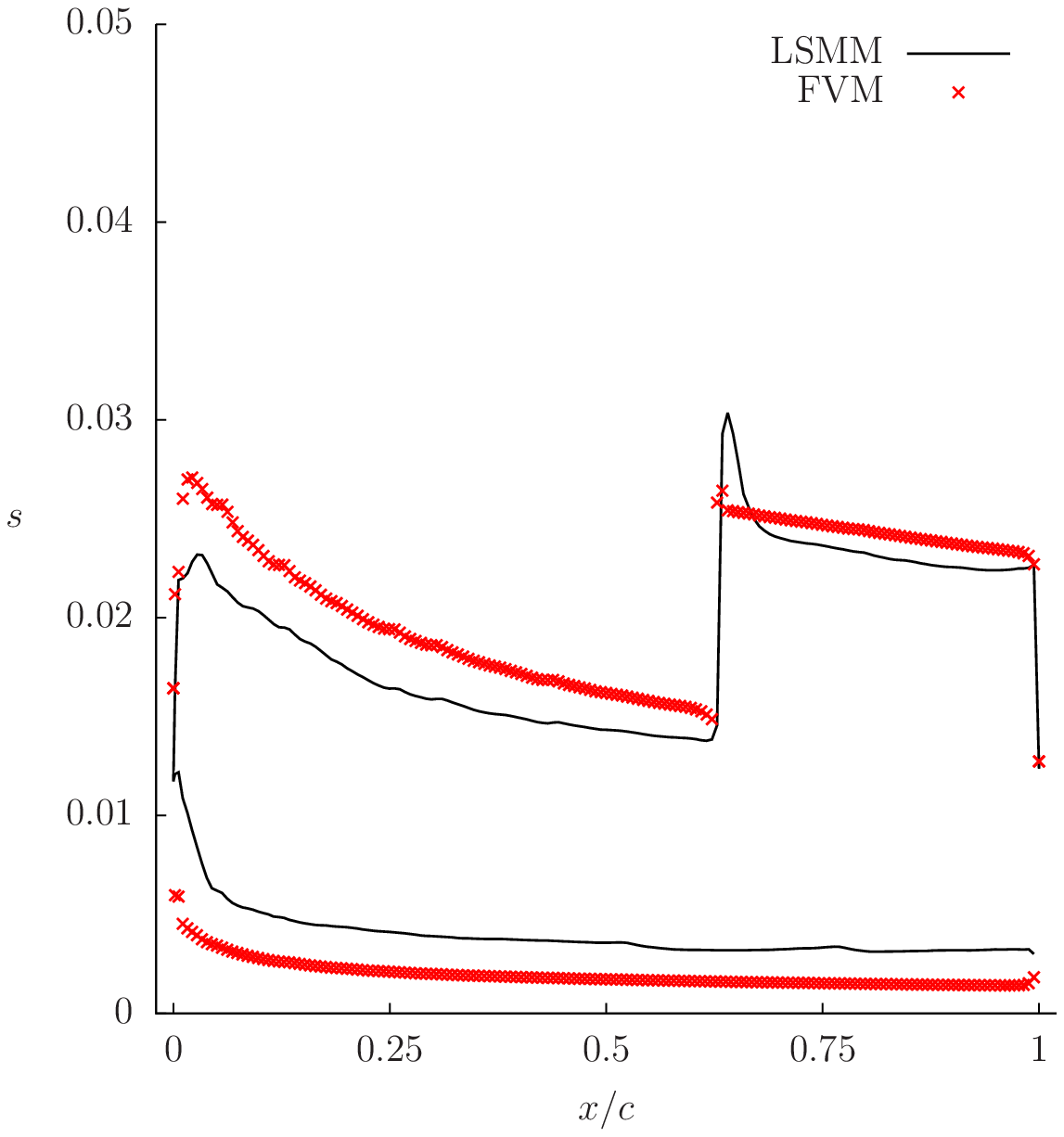}}
  \caption{Pressure coefficient and entropy production around the RAE2822 aerofoil}
  \label{fig:rae2822:cp:s}
\end{figure}

\begin{figure}[htp]
  \centering
  \subfigure[FVM]{
    \includegraphics[width=0.48\textwidth]{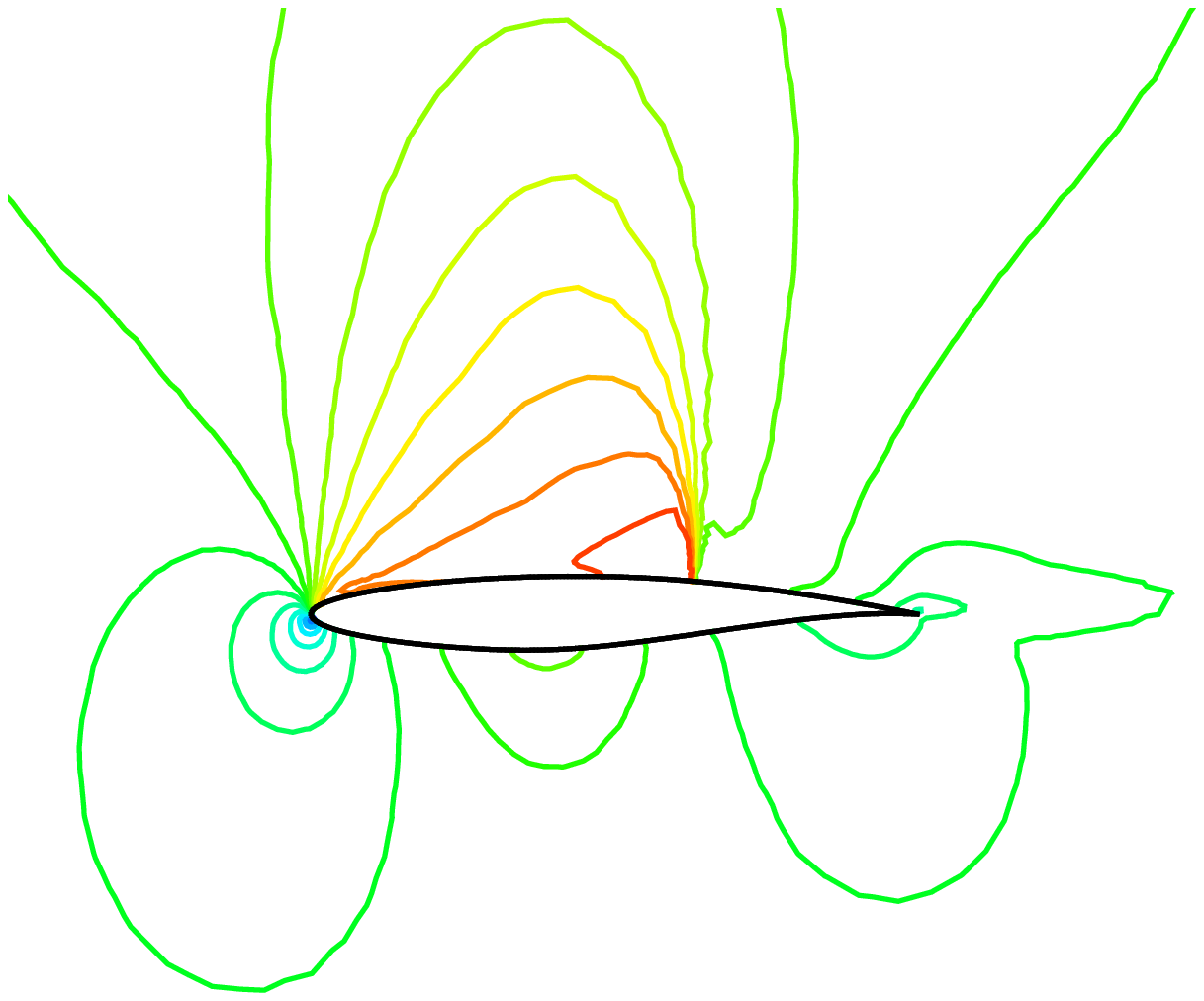}}
  \subfigure[LSMM]{
    \includegraphics[width=0.48\textwidth]{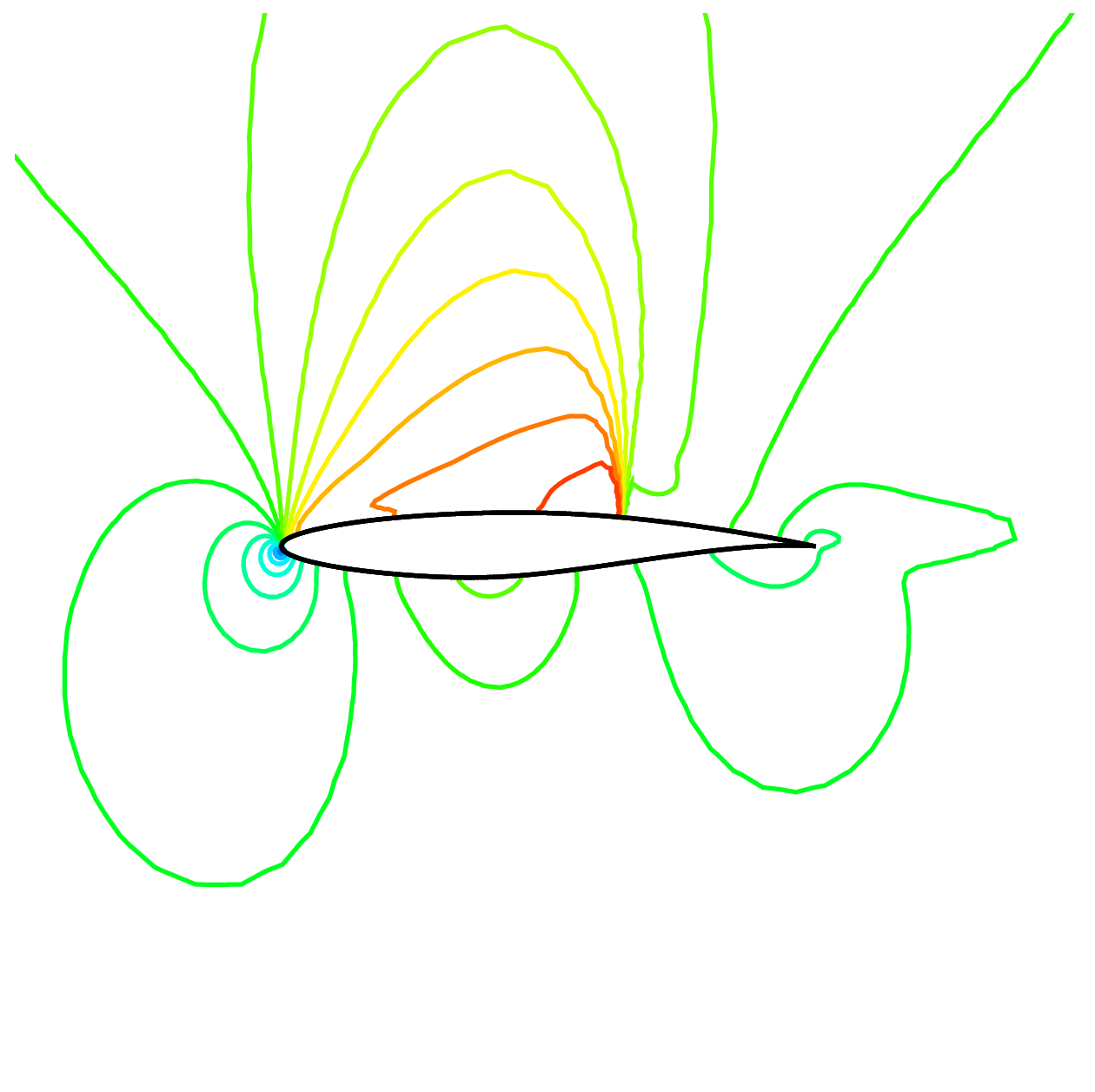}}
  \caption{Mach number contours for the RAE2822 aerofoil}
  \label{fig:rae2822:m073:a280:contour}
\end{figure}

For the RAE2822 aerofoil, we compute the transonic flow with the conditions
$M_{\infty}=0.73$ and $\alpha=2.8^{\rm o}$. The number of meshless points in the
domain is 5,482, there are 42 nodes on the far field boundary and 335 points on
the wall boundary. Figure \ref{fig:rae2822:point} is a snapshot of the scattered
points around the RAE2822 aerofoil. The clouds of points are also constructed by
the natural-neighbour selection criteria. The maximum number of satellite is 9
and the minimum value is 4 in the whole domain. The Mach number contours from
the numerical solutions are shown in Figure \ref{fig:rae2822:m073:a280:contour},
good agreement between the FVM and LSMM results is observed once more. The
pressure coefficients around the aerofoil are shown in Figure
\ref{fig:rae2822:cp}, in which the result computed by a high-order discontinuous
Galerkin method (DGM) from Luo et al. \cite{Paper-2008-Luo-597} is also
included. DGM, FVM and LSMM all underestimate the pressure coefficient at the
leading edge and overestimate at the trailing edge compared to the experiment
\cite{Report-1979-Cook}. The suction peaks captured by DGM, LSMM and FVM on the
aerofoil upper surface adjacent to the leading are close to the laboratory
data. 
DGM, FVM and LSMM get the same shock position, while DGM produces a stronger
shock than FVM and LSMM.  The shock waves captured by DGM, FVM and LSMM are all
stronger than the experimental result and the location is behind the
experiment. This is due to the lack of physical viscosity in the Euler equations
for inviscid flows.  In general, DGM, FVM and LSMM solutions agree reasonably
well with the experimental data in smooth regions.  The entropy productions 
on the aerofoil surface by FVM and LSMM are depicted in Fig. \ref{fig:rae2822:cp:s}

\section{Conclusions}
\label{sec:conclusions}
A meshless method with the HLLC Riemann solver for compressible flows is
presented in this paper. The spatial derivatives of the Euler equations are
discretised by the least-square scheme and the midpoint flux terms are computed
by the appropriate HLLC Riemann solver for LSMM. Crucial information of the
numerical method is derived and exposed in this paper with a step-by-step
instructive computer algorithm provided to readers. Detailed comparisons with
the available exact, finite volume and other reference results for various
numerical test cases validate the methodology for gas flows. The attempt to
handle compressible liquid flows with the proposed method gets unexpected
positive feedback from the one-dimensional liquid shock tube test.  Future work
will include the development of the present method for solving compressible gas
flow problems with moving boundaries and complicated two-/three-dimensional
compressible liquid flows.

\section*{Acknowledgements}
Prof. Eleuterio F. Toro at the University of Trento (Italy) is greatly
appreciated for kindly offering a free download and usage of his NUMERICA
software for hyperbolic conservation laws.  This joint work was partially
supported by the Engineering and Physical Sciences Research Council (EPSRC),
U.K. (grant number EP/J010197/1 and EP/J012793/1), Jyv\"askyl\"a Doctoral
Program in Computing and Mathematical Sciences (COMAS, grant number 21000630)
and the Department of Mathematical Information Technology, University of
Jyv\"askyl\"a, Finland.  The authors are grateful to Prof. Hongquan Chen at
Nanjing University of Aeronautics \& Astronautics (China) for discussing
meshless methods, Prof. N. Balakrishnan at Indian Institute of Science for
providing a very valuable thesis, Prof. Derek Causon at Manchester Metropolitan
University (UK) for discussing HLLC approximate Riemann solvers, Prof. Hong Luo
at North Carolina State University (USA) for providing the initial condition
setup for the moving shock problem.


\end{document}